\documentclass[a4paper,11pt]{article}
\pdfoutput=1 % if your are submitting a pdflatex (i.e. if you have
\usepackage{jcappub} % for details on the use of the package, please
\usepackage[T1]{fontenc} % if needed

\usepackage{graphicx}
\usepackage[utf8]{inputenc}
\usepackage[english]{babel} 
\usepackage{enumitem}
\usepackage{float}
\usepackage{amssymb}
\usepackage{amsmath}
\usepackage{slashed}
\usepackage{color}
\usepackage{mathtools}
\usepackage{multirow}
\usepackage{hyperref}
\usepackage{subfig}
\usepackage{makecell}
\usepackage{cancel}
\usepackage{hhline}
\usepackage[normalem]{ulem}
\def\baredth{\;\overline{\raise1.0pt\hbox{$'$}\hskip-6pt
		\partial}\;}
\def\edth{\;\raise1.0pt\hbox{$'$}\hskip-6pt\partial\;}

\def\setsymbol#1#2{\expandafter\def\csname #1\endcsname{#2}}
\def\getsymbol#1{\csname #1\endcsname}

\def\Planck{\textit{Planck}}

\newbox\tablebox    \newdimen\tablewidth
\def\leaderfil{\leaders\hbox to 5pt{\hss.\hss}\hfil}
\def\tablenote#1 #2\par{\begingroup \parindent=0.8em
    \abovedisplayshortskip=0pt\belowdisplayshortskip=0pt
    \noindent
    $$\hss\vbox{\hsize\tablewidth \hangindent=\parindent \hangafter=1 \noindent
    \hbox to \parindent{$^#1$\hss}\strut#2\strut\par}\hss$$
    \endgroup}

\def\L2{\ifmmode L_2\else $L_2$\fi}

\def\DeltaT{\ifmmode \Delta T\else $\Delta T$\fi}
\def\deltat{\ifmmode \Delta t\else $\Delta t$\fi}
\def\fknee{\ifmmode f_{\rm knee}\else $f_{\rm knee}$\fi}
\def\Fmax{\ifmmode F_{\rm max}\else $F_{\rm max}$\fi}
\def\solar{\ifmmode{\rm M}_{\mathord\odot}\else${\rm M}_{\mathord\odot}$\fi}
\def\Msolar{\ifmmode{\rm M}_{\mathord\odot}\else${\rm M}_{\mathord\odot}$\fi}
\def\Lsolar{\ifmmode{\rm L}_{\mathord\odot}\else${\rm L}_{\mathord\odot}$\fi}
\def\inv{\ifmmode^{-1}\else$^{-1}$\fi}
\def\mo{\ifmmode^{-1}\else$^{-1}$\fi}
\def\sup#1{\ifmmode ^{\rm #1}\else $^{\rm #1}$\fi}
\def\expo#1{\ifmmode \times 10^{#1}\else $\times 10^{#1}$\fi}
\def\,{\thinspace}
\def\lsim{\mathrel{\raise .4ex\hbox{\rlap{$<$}\lower 1.2ex\hbox{$\sim$}}}}
\def\gsim{\mathrel{\raise .4ex\hbox{\rlap{$>$}\lower 1.2ex\hbox{$\sim$}}}}

\def\simprop{\mathrel{\raise .4ex\hbox{\rlap{$\propto$}\lower 1.2ex\hbox{$\sim$}}}}
\def\deg{\ifmmode^\circ\else$^\circ$\fi}
\def\pdeg{\ifmmode $\setbox0=\hbox{$^{\circ}$}\rlap{\hskip.11\wd0 .}$^{\circ}
          \else \setbox0=\hbox{$^{\circ}$}\rlap{\hskip.11\wd0 .}$^{\circ}$\fi}
\def\arcs{\ifmmode {^{\scriptstyle\prime\prime}}
          \else $^{\scriptstyle\prime\prime}$\fi}
\def\arcm{\ifmmode {^{\scriptstyle\prime}}
          \else $^{\scriptstyle\prime}$\fi}
\newdimen\sa  \newdimen\sb
\def\parcs{\sa=.07em \sb=.03em
     \ifmmode \hbox{\rlap{.}}^{\scriptstyle\prime\kern -\sb\prime}\hbox{\kern -\sa}
     \else \rlap{.}$^{\scriptstyle\prime\kern -\sb\prime}$\kern -\sa\fi}
\def\parcm{\sa=.08em \sb=.03em
     \ifmmode \hbox{\rlap{.}\kern\sa}^{\scriptstyle\prime}\hbox{\kern-\sb}
     \else \rlap{.}\kern\sa$^{\scriptstyle\prime}$\kern-\sb\fi}
\def\ra[#1 #2 #3.#4]{#1\sup{h}#2\sup{m}#3\sup{s}\llap.#4}
\def\dec[#1 #2 #3.#4]{#1\deg#2\arcm#3\arcs\llap.#4}
\def\deco[#1 #2 #3]{#1\deg#2\arcm#3\arcs}
\def\rra[#1 #2]{#1\sup{h}#2\sup{m}}

\def\dots{\relax\ifmmode \ldots\else $\ldots$\fi}
\def\WHzsr{\ifmmode $W\,Hz\mo\,sr\mo$\else W\,Hz\mo\,sr\mo\fi}
\def\mHz{\ifmmode $\,mHz$\else \,mHz\fi}
\def\GHz{\ifmmode $\,GHz$\else \,GHz\fi}
\def\mKs{\ifmmode $\,mK\,s$^{1/2}\else \,mK\,s$^{1/2}$\fi}
\def\muKs{\ifmmode \,\mu$K\,s$^{1/2}\else \,$\mu$K\,s$^{1/2}$\fi}
\def\muKRJs{\ifmmode \,\mu$K$_{\rm RJ}$\,s$^{1/2}\else \,$\mu$K$_{\rm RJ}$\,s$^{1/2}$\fi}
\def\muKHz{\ifmmode \,\mu$K\,Hz$^{-1/2}\else \,$\mu$K\,Hz$^{-1/2}$\fi}
\def\MJysr{\ifmmode \,$MJy\,sr\mo$\else \,MJy\,sr\mo\fi}
\def\MJysrmK{\ifmmode \,$MJy\,sr\mo$\,mK$_{\rm CMB}\mo\else \,MJy\,sr\mo\,mK$_{\rm CMB}\mo$\fi}
\def\microns{\ifmmode \,\mu$m$\else \,$\mu$m\fi}

\def\muK{\ifmmode \,\mu$K$\else \,$\mu$\hbox{K}\fi}
\def\microK{\ifmmode \,\mu$K$\else \,$\mu$\hbox{K}\fi}
\def\muW{\ifmmode \,\mu$W$\else \,$\mu$\hbox{W}\fi}
\def\kms{\ifmmode $\,km\,s$^{-1}\else \,km\,s$^{-1}$\fi}
\def\kmsMpc{\ifmmode $\,\kms\,Mpc\mo$\else \,\kms\,Mpc\mo\fi}
\providecommand{\sorthelp}[1]{}

\usepackage[dvipsnames]{xcolor}

\def\NHUNIT{\ifmmode {\rm \,cm^{-2}} \else $\rm \,cm^{-2}$ \fi} % NH units

\def\muKcmb{\ifmmode \,\mu$K$_{\rm CMB}$\else \,$\mu$K$_{\rm CMB}$\fi}
\newcommand{\planck}{\Planck}

\newcommand{\OmegaM}{\ifmmode\Omega_{\rm M}\else $\Omega_{\rm M}$\fi}

\newcommand{\commander}{{\tt Commander}}

\newcommand{\nilc}{{\tt NILC}}   %MB
\newcommand{\sevem}{{\tt SEVEM}} %MB
\newcommand{\smica}{{\tt SMICA}}

\providecommand{\Planck}{\textit{Planck}}
\providecommand{\planck}{\Planck}
\providecommand{\LiteBIRD}{\texttt{LiteBIRD}} %GZ
\providecommand{\LB}{\LiteBIRD} %GZ
\providecommand{\SimonsObservatory}{\texttt{Simons Observatory}} %GZ
\providecommand{\SO}{\SimonsObservatory} %GZ
\providecommand{\CMBSf}{\texttt{CMB-S4}} %GZ
\providecommand{\Sf}{\CMBSf} %GZ

\providecommand{\text}[1]{\rm{#1}}

\providecommand{\muK}{\mu\rm{K}}

\newcommand{\begm}{\begin{pmatrix}}
\newcommand{\enm}{\end{pmatrix}}

\def\pmb#1{\setbox0=\hbox{#1}%
    \kern-.025em\copy0\kern-\wd0
    \kern.05em\copy0\kern-\wd0
    \kern-.025em\raise.0433em\box0}

\def\p2Y{\;_2Y}
\def\m2Y{\;_{-2}Y}
\newcommand{\mksym}[1]{\ifmmode {\rm #1}\else #1\fi}

\providecommand{\text}[1]{\rm{#1}}

\providecommand{\muK}{\mu\rm{K}}

\providecommand{\healpix}{\texttt{HEALPix}}

\newcommand\ba{\begin{eqnarray}}
\newcommand\ea{\end{eqnarray}}
\newcommand\bea{\begin{eqnarray}}
\newcommand\eea{\end{eqnarray}}

\newcommand\be{\begin{equation}}
\newcommand\ee{\end{equation}}

\newcommand{\Nside}{\ensuremath{N_{\mathrm{side}}}}

\title{\boldmath \textit{Planck} constraints on Cosmic Birefringence and its cross-correlation with the CMB}

\author[a,b]{G. Zagatti,}
\author[a,b]{M. Bortolami,}
\author[c,d,a]{A. Gruppuso,}
\author[a,b]{P. Natoli,}
\author[a,b,e]{L. Pagano,}
\author[f,g]{and G. Fabbian}

\affiliation[a]{Dipartimento di Fisica e Scienze della Terra, Università degli Studi di Ferrara, via Saragat 1, I-44122 Ferrara, Italy}
\affiliation[b]{Istituto Nazionale di Fisica Nucleare, Sezione di Ferrara, via Saragat 1, I-44122 Ferrara, Italy}
\affiliation[c]{Istituto Nazionale di Astrofisica - Osservatorio di Astrofisica e Scienza dello Spazio di Bologna, via Gobetti 101, I-40129 Bologna, Italy}
\affiliation[d]{Istituto Nazionale di Fisica Nucleare, Sezione di Bologna, viale Berti Pichat 6/2, I-40127 Bologna, Italy}
\affiliation[e]{Institut d'Astrophysique Spatiale, CNRS, Univ. Paris-Sud, Universit\'{e} Paris-Saclay, B\^{a}t. 121, 91405 Orsay cedex, France}
\affiliation[f]{School of Physics and Astronomy, Cardiff University, The Parade, Cardiff, Wales CF24 3AA, United Kingdom}
\affiliation[g]{Center for Computational Astrophysics, Flatiron Institute, 162 Fifth Avenue, New York, NY, 10010, USA}

\emailAdd{giorgia.zagatti@unife.it}
\emailAdd{marco.bortolami@unife.it}
\emailAdd{alessandro.gruppuso@inaf.it}
\emailAdd{paolo.natoli@unife.it}
\emailAdd{luca.pagano@unife.it}
\emailAdd{fabbiang@cardiff.ac.uk}

\abstract{Cosmic birefringence is the in-vacuo, frequency independent rotation of the polarization plane of linearly polarized radiation, induced by a parity-violating term in the electromagnetic Lagrangian. We implement an harmonic estimator for the birefringence field that only relies on the CMB E to B mode cross-correlation, thus suppressing the effect of cosmic variance from the temperature field. We derive constraints from \planck\ public releases 3 and 4, revealing a cosmic birefringence power spectrum consistent with zero at about $2\sigma$ up to multipole $L=1500$. Moreover, we find that the cross-correlations of cosmic birefringence with the CMB T-, E- and B-fields are also well compatible with null. The latter two cross-correlations are provided here for the first time up to $L=1500$.}

\begin{document}
\maketitle
\flushbottom

\section{Introduction}\label{sec:intro}

The Cosmic Microwave Background (CMB), a radiation that marks the transition from an opaque to a transparent Universe, is a key observable for investigating cosmological physics.
For decades, CMB experiments \cite{Mather:1990tfx,Jones:2005yb,Hanany:2000qf,hinshaw2012} mainly focused on the temperature field of the CMB radiation, whose information was extracted almost completely by the \planck\ satellite \cite{planck2016-l01} up to few arc-minute scale.
Much of the current and future experimental effort is devoted to measuring the polarization part of the CMB radiation \cite{ACT:2020gnv,SPT-3G:2022hvq, BICEP:2021xfz,LiteBIRD:2022cnt,SimonsObservatory:2018koc,Abazajian:2019eic} which is linearly polarized at the $1-10 \%$ level due to Thomson scattering.
The polarization field is usually decomposed into two linear polarization modes: the E-mode, which is parity-even and couples to both scalar and tensor perturbations, and the B-mode, which is parity-odd and exclusively couples to tensor perturbations \cite{Dodelson:2003ft}.
In the standard scenario, the polarization pattern of the CMB is described by the Maxwell's electromagnetism, which preserves parity symmetry. 
In such a case the electromagnetic Lagrangian is described by
\begin{equation}
    \mathcal{L}_{em}^{SM} = -\dfrac{1}{4}F_{\mu\nu}F^{\mu\nu} \, ,
    \label{electromagnetism}
\end{equation}
where $F_{\mu\nu}$ is the electromagnetic tensor that contains the electric and magnetic fields.
Since Eq.~(\ref{electromagnetism}) satisfies parity symmetry, it is possible to show that the CMB cross-correlations TB and EB are expected to be zero.  

However, there are recent claims of deviations from null of the latter cross-correlations \cite{Minami:2020odp,Diego-Palazuelos:2022mcp,Eskilt:2022wav, Eskilt:2022cff}, which are consistent with the Cosmic Birefringence (CB) effect \cite{Carroll:1989vb}, i.e. the rotation of linear polarisation plane of photons during propagation.
Specifically these papers, which are based on \planck\ data and make use of a new technique \cite{Minami:2019ruj,Minami:2020xfg,Minami:2020fin} able to disentangle the instrumental polarization angle from the CB effect\footnote{Otherwise the uncertainty of the instrumental polarization angle has to be assessed independently in the total error budget, see e.g. \cite{Pagano:2009kj,planck2014-a23}.}, hint at a detection of a CB angle $\beta \sim 0.3^{\circ}$ at the level of 2.5 - 3$\sigma$.
The latter CB angle is also called \textit{isotropic}, meaning that it does not depend on the direction of observation. 
There is also an \textit{anisotropic} CB effect, which instead depends on the direction of observations, that is currently found to be well compatible with null \cite{Contreras:2017sgi,SPT:2020cxx,Namikawa:2020ffr,Gruppuso:2020kfy,Bortolami:2022whx}. See also \cite{Komatsu:2022nvu} for a review of the CB effect from CMB observations.

If not due to systematic effects of instrumental or astrophysical origin \cite{Clark:2021kze,Cukierman:2022kei,Diego-Palazuelos:2022cnh}, these analyses hinting at an isotropic CB effect, suggest the need to extend the electromagnetic sector of the standard model $\mathcal{L}_{em}^{SM}$ with a parity-violating term $\mathcal{L}_{CS}$
\begin{equation}
    \mathcal{L}_{em} = \mathcal{L}_{em}^{SM} + \mathcal{L}_{CS} = -\dfrac{1}{4}F_{\mu\nu}F^{\mu\nu} -\dfrac{\lambda}{4 f} \phi \, F_{\mu\nu} \Tilde{F}^{\mu\nu} ,
    \label{electromagnetismext}
\end{equation}
known as a Chern-Simons term \cite{Chern:1974ft}\footnote{For other extensions, see e.g. \cite{Gubitosi:2009eu, Lembo:2020ufn}}.
In Eq.~(\ref{electromagnetismext}), $\lambda/f$ is a coupling with the dimension of the inverse of an energy, $\phi$ is a new scalar (or pseudo-scalar) field and $\Tilde{F}^{\mu\nu}$ is the dual electromagnetic tensor. %\cite{Komatsu:2022nvu,PhysRevD.43.3789}.
With such an extension, it is possible to describe an isotropic CB effect, and consequently a CMB EB cross-correlation compatible with observations, when $\phi$ is taken to be homogeneous
\cite{Murai:2022zur,Eskilt:2023nxm}.
Fluctuations of $\phi$ around its homogeneous part, are instead able to produce anisotropic CB \cite{Li:2008tma, Kamionkowski:2008fp}.
Hence, the CB effect can be seen as a tracer of the existence of a new cosmological field $\phi$ (typically referred as an axion) acting as dark matter or dark energy \cite{Carroll:1998zi,Panda:2010uq,Marsh:2015xka,Ferreira:2020fam} which might also play a role in alleviating the Hubble tension, see e.g. \cite{Capparelli:2019rtn,DiValentino:2021izs} and reference therein. See also \cite{Finelli:2008jv,Fedderke:2019ajk,Nakatsuka:2022epj,Galaverni:2023zhv,Greco:2022xwj}.

Other works, e.g. \cite{Eskilt:2023nxm,Greco:2024oie}, have constrained the axion's parameters through the CMB EB\footnote{Current detections of the isotropic CB effect are based on the CMB cross-correlation EB and not TB, since the latter has a lower signal to noise ratio, at least a factor of 2 for \planck\ \cite{planck2014-a23}.} power spectrum or the isotropic CB effect. 
However, these models can be put on additional tests considering also the anisotropic CB and the cross-correlations between the anisotropic CB and the CMB field, see e.g. \cite{Caldwell:2011pu}. 
Therefore it is essential to provide updated constraints on the latter observables. %which can be confronted with the models. 
For this reason in this work, we focus on the anisotropic CB effect implementing an harmonic estimator based on the approach presented in \cite{Gluscevic:2009mm} and \cite{Namikawa:2020ffr}. 
The aim is to apply this estimator to the most recent \planck\,data, namely PR3 \cite{planck2016-l01} and PR4 (also known as NPIPE) \cite{planck2020-LVII}, and to provide new constraints on the CB power spectrum and the CB cross-correlation with the CMB fields. 
In particular the cross-correlation between anisotropic CB and the CMB E- and B-fields are given here for the first time up to $L=1500$ (previously in \cite{Bortolami:2022whx}, considering a different technique, those were provided only at very low multipoles). \par
This paper is organized as follows. In sections \ref{sec:harmonic estimator} and \ref{sec:debias}, we present the methodology employed to estimate the CB power spectrum, describing the structure of the estimator for the spherical harmonic coefficients and the de-biasing procedure necessary to obtain the final estimate. In section \ref{sec:data_sim}, we describe the CMB data and simulation sets utilized in this study. In section \ref{sec:Results}, we present the results of applying our pipeline to \planck\,CMB polarization maps and the cross-correlations of the CB and the CMB temperature and polarization fields. Furthermore, in section \ref{sec:forecasts}, we forecast the sensitivity of forthcoming CMB experiments, such as the \LB\ satellite, the \SO, and \Sf\ to the EB cross-correlation. We conclude in section \ref{sec:conclusions}. The full calculations leading to the final expression of the estimator can be found in appendix \ref{app:impl est}, and validation tests for our pipeline are presented in appendix \ref{app:validation}.

\section{Harmonic Estimator} \label{sec:harmonic estimator}
This section introduces the impact of CB on CMB observations, in order to provide the structure of the harmonic estimator used in this work.\par
The primary effect is that the \textit{observed} CMB polarization field carries the information of the rotation field \cite{Kamionkowski:2008fp}. Consequently, we do not observe the primordial E-modes and B-modes, denoted as $a_{\ell m}^X$, instead we observe their sum with the rotation-induced modes, $\delta a_{\ell m}^X$. Here $X=E,B$ for E- and B-modes respectively:
\begin{align}\label{rot_E}
    &a_{\ell m}^{E, tot} = a_{\ell m}^E + \delta a_{\ell m}^E , \\ \label{rot_B}
    &a_{\ell m}^{B, tot} = a_{\ell m}^B + \delta a_{\ell m}^B \simeq \delta a_{\ell m}^B , 
\end{align}
where the second equivalence in equation \eqref{rot_B} holds since we are assuming that the B-modes generated on the last scattering surface are null\footnote{In this study we present the calculations and the results based on this assumption. However, the entire analysis can be generalized to the case of non-zero B-modes on the last scattering surface}. The expressions for the rotation-induced E-modes and B-modes, as derived in \cite{Kamionkowski:2008fp}, are given as:
\begin{align}\label{delta_B}
    &\delta a_{\ell m}^B = 2\sum_{LM}\sum_{\ell'm'}\alpha_{LM}a_{\ell'm'}^E\xi_{\ell m\ell'm'}^{LM}H_{\ell\ell'}^L, \\ \label{delta_E}
    &\delta a_{\ell m}^E = 2i\sum_{LM}\sum_{\ell'm'}\alpha_{LM}a_{\ell'm'}^E\xi_{\ell m\ell'm'}^{LM}H_{\ell\ell'}^L,  
\end{align}
where equation \eqref{delta_B} is different from zero for $\ell + \ell' + L$ even, equation \eqref{delta_E} is different from zero for $\ell + \ell' + L$ odd, $\alpha_{LM}$ are the spherical harmonic coefficients of the CB field, and $\xi_{\ell m\ell'm'}^{LM}$ and $H_{\ell\ell'}^L$ are defined in terms of Wigner-3j symbols as follows:
\begin{align}
    &\xi_{\ell m\ell'm'}^{LM} = (-1)^{m}\sqrt{\dfrac{(2\ell+1)(2L+1)(2\ell'+1)}{4\pi}}\begin{pmatrix}
        \ell & L & \ell' \\
        -m & M & m'
    \end{pmatrix} , \\
    &H_{\ell\ell'}^L = \begin{pmatrix}
        \ell & L & \ell' \\
        2 & 0 & -2
    \end{pmatrix} .
\end{align}
The rotation-induced modes generate correlations between $\ell-\ell'$ pairs with $\ell \neq \ell'$, leading to, as anticipated in section \ref{sec:intro}, parity-violating cross-correlations. These non-standard cross-correlations caused by CB can be characterized by the following general structure:
\begin{equation}\label{cross-corr}
    \bigl<a_{\ell m}^{X, tot}a_{\ell'm'}^{X',tot,*}\bigr> = 2\displaystyle\sum_{LM}\alpha_{LM}Z_{\ell\ell'}^{XX'}\xi_{\ell m \ell'm'}^{LM}H_{\ell\ell'}^L ,
\end{equation}
where $X=\{T,E,B\}$ and $Z_{\ell\ell'}^{XX'}$ contains the information about the primordial spectra before the rotation (see table 1 of \cite{Gluscevic:2009mm}). Another way to write equation \eqref{cross-corr} is in terms of the \textit{rotational invariants} (\textit{i.e.} quantities independent of \textit{m}) \cite{Kamionkowski:2008fp}, $D_{\ell\ell'}^{LM,XX'}$:
\begin{equation}
    \bigl<a_{\ell m}^{X, tot}a_{\ell'm'}^{X',tot,*}\bigr> = \displaystyle\sum_{LM}D_{\ell\ell'}^{LM,XX'}\xi_{\ell m \ell'm'}^{LM} ,
\end{equation}
where,
\begin{equation}\label{rot_inv}
    D_{\ell\ell'}^{LM,XX'} = 2\alpha_{LM}Z_{\ell\ell'}^{XX'}H_{\ell\ell'}^L .
\end{equation}
The starting point to get an expression for the estimator are the rotational invariants, indeed. Their definition in equation \eqref{rot_inv} refers to the primordial signal, thus when moving to observations we have to account for the window function, approximated with a Gaussian symmetric beam, $W_\ell$, so that \cite{Gluscevic:2009mm}:
\begin{equation}\label{obs_rot_inv}
    D_{\ell\ell'}^{LM,XX',map} = D_{\ell\ell'}^{LM,XX'} W_{\ell}W_{\ell'} = 2\alpha_{LM}Z_{\ell\ell'}^{XX'}H_{\ell\ell'}^L W_{\ell}W_{\ell'} ,
\end{equation}
where we use the superscript \textit{``map''} to denote quantities recovered from CMB maps.\par
Having the analytical definition of the observed rotational invariants, we can now provide for two different expressions of the associated estimators. In this work we use an over-hat symbol to indicate \textit{estimated} quantities. The first expression is directly derived from equation \eqref{obs_rot_inv}:
\begin{equation}\label{est_rot_1}
    \hat{D}_{\ell\ell'}^{LM,XX',map} = 2\hat{\alpha}_{LM}Z_{\ell\ell'}^{XX'}H_{\ell\ell'}^L W_{\ell}W_{\ell'} ,
\end{equation}
and the other is the inverse variance weighting average estimator from \cite{Pullen:2007tu}:
\begin{equation}\label{est_rot_2}
    \hat{D}_{\ell\ell'}^{LM,XX',map} = (G_{\ell\ell'})^{-1}\displaystyle\sum_{mm'}a_{\ell m}^{X,map}a_{\ell'm'}^{X',map,*}\xi_{\ell m \ell'm'}^{LM} ,
\end{equation}
where $G_{\ell\ell'}$ is defined as:
\begin{equation}
    G_{\ell\ell'} = \displaystyle\sum_{mm'}(\xi_{\ell m\ell'm}^{LM})^2 = \dfrac{(2\ell+1)(2\ell'+1)}{4\pi} .
\end{equation}
The first expression of the estimates for the $\alpha_{LM}$ coefficients for a fixed $\ell-\ell'$ pair is obtained inverting equation \eqref{est_rot_1} and substituting the expression of the estimator for the rotational invariants of equation \eqref{est_rot_2}: 
\begin{equation}\label{alpha}
    (\overline{\alpha}_{LM})_{\ell\ell'}^{XX'} = \dfrac{\hat{D}_{\ell\ell'}^{LM,XX',map}}{F_{\ell\ell'}^{L,XX'}} = \dfrac{(G_{\ell\ell'})^{-1}\displaystyle\sum_{mm'}a_{\ell m}^{X,map}a_{\ell'm'}^{X',map,*}\xi_{\ell m \ell'm'}^{LM}}{F_{\ell\ell'}^{L,XX'}},
\end{equation}
where we defined $F_{\ell\ell'}^{L,XX'} = 2Z_{\ell\ell'}^{XX'}H_{\ell\ell'}^{L}W_\ell W_{\ell'}$ \cite{Kamionkowski:2008fp}. Note that we are not using the over-hat symbol since, before ending up with the final estimates of the $\alpha_{LM}$ coefficients, we have to encode for a de-biasing procedure.\par
The final expression of the harmonic estimator before the de-biasing procedure at the level of the spherical harmonic coefficients, has been obtained applying the definition of the \textit{inverse variance weighting average} to the $(\overline{\alpha}_{LM})_{\ell\ell'}^{XX^\prime}$ defined in equation \eqref{alpha}, ending up with:
\begin{equation}\label{min_var_est}
    \overline{\alpha}_{LM}^{XX'} = \dfrac{\displaystyle\sum_{\ell\ell'}(\overline{\alpha}_{LM})_{\ell\ell'}^{XX'}/(\sigma_L^2)_{\ell\ell'}^{XX'}}{\displaystyle\sum_{\ell\ell'}1/(\sigma_L^2)_{\ell\ell'}^{XX'}} ,
\end{equation}
where $(\sigma_L^2)_{\ell\ell'}^{XX'}$ is the \textit{analytic} variance associated to each $(\overline{\alpha}_{LM})_{\ell\ell'}$ estimate and can be obtained after a proper re-scaling of the variance of the rotational invariants (see eq. \eqref{est_rot_1} for the re-scaling):
\begin{equation}\label{sigma_teo}
    (\sigma_L^2)_{\ell\ell'}^A = \dfrac{\mathcal{C}_{AA'}^{\ell\ell'}}{G_{\ell\ell'}F_{\ell\ell'}^{L,A}F_{\ell\ell'}^{L,A'}} ,
\end{equation}
where $A$ represents the considered cross-correlation, \textit{i.e.} $XX'$, and $\mathcal{C}_{AA'}^{\ell\ell'}$ is the entry $AA'$ of the covariance matrix of the rotational invariants (see equation 33 of \cite{Gluscevic:2009mm} for the full expression of the covariance matrix).\par
In this study, we employ the analytic variance to normalize the estimator. While it is important to note that the analytic variance is rigorously justified under the conditions of full-sky observations with homogeneous noise, our investigation has demonstrated that even for cut-sky observations, the analytic expression approximates the true variance of the estimator effectively.\par
After obtaining the initial estimates for the $\alpha_{LM}$ coefficients, we followed the approach outlined in \cite{SPT:2020cxx}, which encodes for a de-biasing procedure to derive the final estimates of the spherical harmonic coefficients of the CB field. Having the de-biased $\alpha_{LM}$ coefficients allows us to evaluate the CB power spectrum (section \ref{sec:power spectrum}) and the map of the CB field (section \ref{subsec:cross_corr})\par
To end up with the un-biased estimates of the $\alpha_{LM}$ we have to subtract the \textit{mean field} bias. The mean field is a contribution, at the level of maps, coming from mask effects, not homogeneous noise, and other signals of the map different from the CB field.\par
The evaluation of the mean field bias entirely relies on simulations:
\begin{equation}\label{mf bias}
    \alpha_{LM}^{bias, MF} = <\overline{\alpha}_{LM}>_{sim} ,
\end{equation}
where the $\overline{\alpha}_{LM}$ are the spherical harmonic coefficients of the CB field evaluated over the simulations of the CMB maps from equation \eqref{alpha}, and averaged over the entire simulation set.\par
Thus, the final estimates of the $\alpha_{LM}$ coefficients have been obtained as:
\begin{equation}\label{de-biased alms}
    \hat{\alpha}_{LM} = \overline{\alpha}_{LM} - \alpha_{LM}^{bias,MF} .
\end{equation}
Next, we will focus specifically on the EB cross-correlation.

\subsection{EB-estimator}\label{subsec:eb-estimator}
In the following, we are going to focus on the EB cross-correlation only, showing how the information contained in the CMB polarization field can be used to develop an estimator for the spherical harmonic coefficients of the CB field.\par 
The EB cross-correlation is induced by a rotation of the primordial EE power spectrum, while the TB cross-correlation is generated from the rotation of the primordial TE power spectrum. Observations involving the CMB temperature field are affected by the cosmic variance. For the \planck\,satellite, as well as for the forthcoming CMB experiments, the signal-to-noise ratio for the EB signal is larger than the one for TB. For this reason, in this study we implement the estimator based on the information coming from the EB signal only.\par
Furthermore, in section \ref{sec:debias}, starting from the estimates of the spherical harmonic coefficients of the rotation field, we present the procedure to estimate the CB power spectrum.\par
The EB cross-correlation, assuming that B-modes on the last scattering surface are null, is:
\begin{align}\label{EB spectrum}
    \bigl<a_{\ell m}^{E,tot} a_{\ell'm'}^{B,tot,*}\bigr> &= \bigl<(a_{\ell m}^E + \delta a_{\ell m}^E) \delta a_{\ell' m'}^{B,*}\bigr> = \nonumber \\
        &= \bigl<a_{\ell m}^E \delta a_{\ell'm'}^{B,*}\bigr> + \bigl<\delta a_{\ell m}^E \delta a_{\ell' m'}^{B,*}\bigr> \simeq  \bigl<a_{\ell m}^E \delta a_{\ell'm'}^{B,*}\bigr> , 
\end{align}
where the last equivalence holds since the term $\bigl<\delta a_{\ell m}^E \delta a_{\ell' m'}^{B,*}\bigr>$ is second order and we neglect it. From equation \eqref{EB spectrum} we see that the EB cross-correlation is determined by the rotated B-modes, thus it is different from zero if $\ell + \ell' + L$ is even (see equation \eqref{delta_B}).\par
Equation \eqref{cross-corr} for the specific case of the EB cross-correlation now reads:
\begin{equation}\label{EB cross-corr}
    \bigl<a_{\ell m}^{E,tot}a_{\ell'm'}^{B,tot,*}\bigr> = 2\sum_{LM}\alpha_{LM}C_\ell^{EE}\xi_{\ell m\ell'm'}^{LM}H_{\ell\ell'}^L,
\end{equation}
where $C_\ell^{EE}$ refers to the $Z_{\ell\ell'}^{XX'}$ term (eq. \eqref{cross-corr}) when $XX'=EB$.\par
Following the logical steps previously described, the first expression of the estimator for the $\alpha_{LM}$ coefficients before the subtraction of the mean field bias (eq. \eqref{min_var_est}) is calculated as:
\begin{equation}\label{min var EB}
    \overline{\alpha}_{LM}^{EB} = \dfrac{\displaystyle\sum_{\ell\ell'}(\overline{\alpha}_{LM})_{\ell\ell'}^{EB}/(\sigma_L^2)_{\ell\ell'}^{EB}}{\displaystyle\sum_{\ell\ell'}1/(\sigma_L^2)_{\ell\ell'}^{EB}} ,
\end{equation}
and the expression for the inverse variance of the estimator for the EB cross-correlation is:
\begin{equation}\label{sigma}
    \sigma_L^{-2} = \displaystyle\sum_{\ell\geq\ell'}(1+\delta_{\ell\ell'})^{-1}G_{\ell\ell'}\Biggl\{\dfrac{(F_{\ell\ell'}^{L,EB})^2}{C_\ell^{EE,map}C_{\ell'}^{BB,map}} + \dfrac{(F_{\ell\ell'}^{L,BE})^2}{C_\ell^{BB,map}C_{\ell'}^{EE,map}}\Biggr\} .
\end{equation}
Focusing only on the numerator in equation \eqref{min var EB} (meaning the unnormalized estimator, \textit{UN}):
\begin{equation}\label{T_int}
    \overline{\alpha}_{LM}^{UN} = \displaystyle\sum_{\ell\ell'}\left\{\dfrac{F_{\ell\ell'}^{L,EB}\displaystyle\sum_{mm'}a_{\ell m}^{E,map}a_{\ell'm'}^{B,map,*}\xi_{\ell m\ell'm'}^{LM}}{C_\ell^{EE,map}C_{\ell'}^{BB,map}}\right\} ,
\end{equation}
and exploiting properties of Wigner-3j symbols, it is possible to re-write equation \eqref{T_int} as\footnote{Details about the calculations are provided in appendix \ref{app:impl est}.}:
\begin{multline}\label{new_est}
\overline{\alpha}_{LM}^{UN} =
\int d\hat{n} Y_{LM} \Biggl[\displaystyle\sum_{\ell m}\dfrac{C_\ell^{EE}a_{\ell m}^{E,map,*}W_\ell {}_{-2}Y_{\ell m}}{C_\ell^{EE,map}}\displaystyle\sum_{\ell'm'}\dfrac{a_{\ell'm'}^{B,map,*}W_{\ell'}{}_{+2}Y_{\ell'm'}}{C_{\ell'}^{BB,map}}+\\
\displaystyle\sum_{\ell m}\dfrac{C_\ell^{EE}a_{\ell m}^{E,map,*}W_\ell {}_{+2}Y_{\ell m}}{C_\ell^{EE,map}}\displaystyle\sum_{\ell'm'}\dfrac{a_{\ell'm'}^{B,map,*}W_{\ell'}{}_{-2}Y_{\ell'm'}}{C_{\ell'}^{BB,map}}\Biggr] ,
\end{multline}
where ${}_{\pm 2}Y_{\ell m}$ are spin-2 spherical harmonics and $C_\ell^{XX,map}$ indicates the analytical expression for the power spectrum recovered from CMB maps, evaluated as:
\begin{equation}\label{cl_teo}
    C_\ell^{XX,map} = C_\ell^{XX}W^2_\ell + N_\ell^{XX} ,
\end{equation}
with $C_\ell^{XX}$ the cosmological signal, $W_\ell$ the \textit{window function}, and $N_\ell^{XX}$ the noise power spectrum. Note that we are using the superscript \textit{UN} to indicate that we refer to the first estimate of the estimator (eq. \ref{min var EB}) without its normalization (eq. \eqref{sigma}).\par
Following the approach of \cite{Namikawa:2020ffr}, it is possible to re-write the expression of the estimator in a way so that the computational time is remarkably reduced, defining the two following new objects:
\begin{align}\label{QUobjects_E}
    Q^E \pm iU^E = \displaystyle\sum_{\ell m}(C_\ell^{EE}\overline{a}_{\ell m}^{E,*}){}_{\pm 2}Y_{\ell m}, \\\label{QUobjects_B}
    Q^B \pm iU^B = \displaystyle\sum_{\ell m}(\pm i\overline{a}_{\ell m}^{B,*}){}_{\pm 2}Y_{\ell m} ,
\end{align}
with $\overline{a}_{\ell m}^{E,*}$ and $\overline{a}_{\ell m}^{B,*}$ defined as:
\begin{align}
    \overline{a}_{\ell m}^{E,*} = \dfrac{a_{\ell m}^{E,map,*}}{C_\ell^{EE,map}}W_\ell , \\
    \overline{a}_{\ell m}^{B,*} = \dfrac{a_{\ell m}^{B,map,*}}{C_\ell^{BB,map}}W_\ell .
\end{align}
By making use of equations \eqref{QUobjects_E} and \eqref{QUobjects_B}, the unnormalized estimator for the $\alpha_{LM}$ coefficients can be written as:
\begin{equation}\label{alm_map}
    \overline{\alpha}_{LM}^{UN} = \int d\hat{n} Y_{LM}[2(Q^EU^B-U^EQ^B)].
\end{equation}
At this point we define the quantity inside the square brackets as a \textit{``map-like''} object:
\begin{equation}\label{new_map}
    m'(\alpha) = 2(Q^E U^B - U^E Q^B) ,
\end{equation}
and, after performing the complex conjugate of equation \eqref{alm_map}, we obtain:
\begin{equation}\label{coeff_cc}
    \overline{\alpha}_{LM}^{UN,*} = \int d\hat{n}Y_{LM}^* m'^*(\alpha) = \int d\hat{n}Y_{LM}^* m'(\alpha), 
\end{equation}
where the second equivalence holds since maps are real objects, \textit{i.e.} $m'^*(\alpha) = m'(\alpha)$. We need to perform the complex conjugate of equation \eqref{alm_map} since otherwise we do not have the correct relation that allows to move from the map to the spherical harmonic coefficients.\par
The above equation is what allows the reduction of the computational time since, having defined the map-like object $m'(\alpha)$ in eq. \eqref{new_map} the computation of the associated spherical harmonic coefficients is straightforward.\\
A word of caution before proceeding. Equation \eqref{coeff_cc} provides for the \textit{complex conjugate} of the unnormalized $\alpha_{LM}$ estimates, thus the estimates of the $\alpha_{LM}$ coefficients before the de-biasing procedure are obtained as:
\begin{equation}\label{alpha_est}
    \overline{\alpha}_{LM} = \dfrac{(\overline{\alpha}_{LM}^{UN,*})^*}{\sigma_L^{-2}} .
\end{equation}
To end up with the final estimates of the spherical harmonic coefficients of the CB field, \textit{i.e.} $\hat{\alpha}_{LM}$, we have to subtract the mean field bias (eq. \eqref{mf bias}) from equation \eqref{alpha_est}, as presented in equation \eqref{de-biased alms}.

\section{Angular power spectrum and de-biasing procedure}\label{sec:power spectrum}
\label{sec:debias}
Having the estimates of the $\alpha_{LM}$ coefficients, the $\alpha\alpha$ power spectrum can be written as:
\begin{equation}\label{cl_aa_map}
    C_L^{\hat{\alpha}\hat{\alpha}} = \dfrac{1}{f_{sky}}\dfrac{1}{2L+1}\displaystyle\sum_M\hat{\alpha}_{LM}\hat{\alpha}_{LM}^* ,
\end{equation}
where $f_{sky}$ is the sky fraction of the mask used for the analysis.\par
The estimate of the CB power spectrum in equation \eqref{cl_aa_map} is intrinsically biased, that is, $C_L^{\hat{\alpha}\hat{\alpha}} \neq \hat{C}_L^{\alpha\alpha}$. The bias comes from the diagonal contribution of $\ell-\ell'$ pairs with $\ell = \ell'$ when combining together the two estimates of the $\alpha_{LM}$ coefficients and from the off-diagonal contribution from sources different from the rotation induced by CB.  We thus encode for a de-biasing procedure \cite{Gluscevic:2012me}, now at the level of the power spectrum, to, first, select the diagonal contribution (\textit{i.e.} the contribution from $\ell-\ell'$ pairs with $\ell = \ell'$) and, second, among the off-diagonal contributions, select the contribution coming \textit{only} from the rotation induced by CB:
\begin{equation}\label{PS_est}
    \hat{C}_L^{\alpha\alpha} = C_L^{\hat{\alpha}\hat{\alpha}} - C_L^{bias} .
\end{equation}
The bias term accounts for two different contributions:
\begin{itemize}
    \item the \textit{isotropic bias term}, $C_L^{bias,iso}$, which is an analytic bias term calculated on data, so that it selects the diagonal contribution coming from the $\ell-\ell'$ pairs with $\ell=\ell'$;
    \item the \textit{Monte Carlo bias term}, $C_L^{bias,MC}$. This term is based on Monte Carlo simulations, in order to describe the off-diagonal signal coming from other contributions different from CB, such as not homogeneous noise, cut-sky effects and the contribution coming from lensing. At the level of the spherical harmonics coefficients of the CB field, lensing has no contribution since CB and lensing are orthogonal effects (\cite{2009PhRvD..80b3510G}).  However, at the power spectrum level, this assertion does not hold true. Therefore, we include a bias term to account for and subtract the contribution of lensing.
\end{itemize}
For this specific case, meaning for the application of the pipeline to \planck\,data, the de-biasing procedure to obtain the final estimate of the power spectrum can entirely rely on simulations, without the need of the analytic computation of the bias. Despite that, in this work we present the general de-biasing procedure that encodes for the analytic bias term too.\par
The isotropic bias term is defined as:
\begin{equation}\label{bias_iso_theo}
    C_L^{bias,iso} = <\hat{\alpha}_{LM}\hat{\alpha}_{LM}^*> .
\end{equation}
From a general point of view, the estimator (eq.\eqref{T_int}) probes three disconnected Wick contractions \cite{Gluscevic:2012me}: EE-BB, EB-EB, and EB-BE.
The cross-correlation terms (\textit{i.e.} EB-EB and EB-BE) are negligible with respect to the auto-correlation terms (\textit{i.e.} EE-BB), both with and without a rotation signal induced by CB. Under this assumption, the equation for the isotropic bias term reduces to:
\begin{eqnarray} \label{bias_iso}
    C_L^{bias,iso} = \dfrac{1}{\sigma_L^{-2}\sigma_L^{-2}}\displaystyle\sum_{\ell'\geq\ell}(1+\delta_{\ell\ell'})^{-1}G_{\ell\ell'}&\left\{\dfrac{(F_{\ell\ell'}^{L,EB})^2\hat{C}_\ell^{EE,map}\hat{C}_{\ell'}^{BB,map}}{C_{\ell'}^{BB,map}C_\ell^{EE,map}C_{\ell'}^{BB,map}C_\ell^{EE,map}} +\right.\nonumber\\
    &\left.+\dfrac{(F_{\ell\ell}^{L,BE})^2\hat{C}_{\ell'}^{EE,map}\hat{C}_\ell^{BB,map}}{C_\ell^{BB,map}C_{\ell'}^{EE,map}C_\ell^{BB,map}C_{\ell'}^{EE,map}}\right\} ,
\end{eqnarray}
where $\hat{C}_\ell^{XX,map}$ with $XX=\{EE,BB\}$ are the power spectra estimated from the CMB polarization maps, and corrected by the $1/f_{sky}$ of the applied mask (see section \ref{sec:data_sim} for details about the masks employed in this analysis). Instead, we use the $C_\ell ^{XX,map}$ (without the over-hat symbol) for the analytic expression of the power spectrum including both the cosmological signal and the noise contributions (see equation \eqref{cl_teo}).\par
The Monte Carlo bias term is based on simulations with the aim of evaluating the off-diagonal signal induced by not homogeneous noise, cut-sky effects and lensing. For this reason, the $C_L^{bias, MC}$ is computed over a set of simulations that resemble the data, masked with the fiducial analysis mask used for data themselves, and which do not contain a rotation signal induced by CB:
\begin{equation}\label{bias_MC}
    C_L^{bias,MC} = <C_L^{\hat{\alpha}\hat{\alpha}} - C_L^{bias,iso}>_{sims} .
\end{equation}
Here the brackets indicate the average computed over the simulations. Since, by construction, all the simulations do not have off-diagonal contributions coming from CB and the diagonal contribution is erased by the isotropic bias term for each simulation, the only off-diagonal signal has to come from the correlations induced by not homogeneous noise, cut-sky effects and lensing.\par
The full expression for the bias term in equation \eqref{PS_est} has to encode for both the isotropic and Monte Carlo bias terms (eqs. \eqref{bias_iso} and \eqref{bias_MC}):
\begin{equation}
    C_L^{bias} = C_L^{bias,iso} + C_L^{bias,MC} .
\end{equation}
This methodology has been validated and the results of this validation are reported in appendix \ref{app:validation}. In the following sections we present the data set used for the analysis and the application of the aforementioned pipeline to \planck\,data.\par

\section{Data set and simulations}\label{sec:data_sim}
In this section we describe the data products employed during this work. The results that will be presented in the following have been obtained on the Public Release 3 (PR3) \cite{planck2016-l01} and the Public Release 4 (NPIPE) \cite{planck2020-LVII} data products of the \planck\,satellite, which contain CMB data and simulation maps at the \texttt{HEALPix}\footnote{http://healpix.sourceforge.net}~\cite{gorski2005} resolution of \Nside = 2048. The CMB maps employed for the main results have been cleaned using the official \planck\,component separation method \commander\,\cite{planck2016-l04}. However, we also present a comparison between the different component separation methods as \sevem, \smica\,and \nilc.\par
\planck\,NPIPE has 400 CMB polarization+noise simulations and 100 CMB temperature+noise simulations for the component separation method \commander, and 600 CMB+noise simulations for the component separation method \sevem; all for \textit{full mission} data and for the two data splits, A and B.\par
The available simulations in PR3 are CMB-only and noise simulations. The former accounts for 1000 CMB Monte Carlo simulations obtained using the \planck\,$\Lambda CDM$ best-fit model. For what concerns noise, there are 300 noise simulations for the \textit{half-mission 1}, 300 for the \textit{half-mission 2} and 300 for the \textit{full mission}. The simulation set used in this work is divided in three subsets, obtained adding the CMB Monte Carlo simulations and the 300 noise simulations for the two half missions and for the full mission.\par
\begin{figure}[ht]  
\centering
    \includegraphics[width=.32\linewidth]{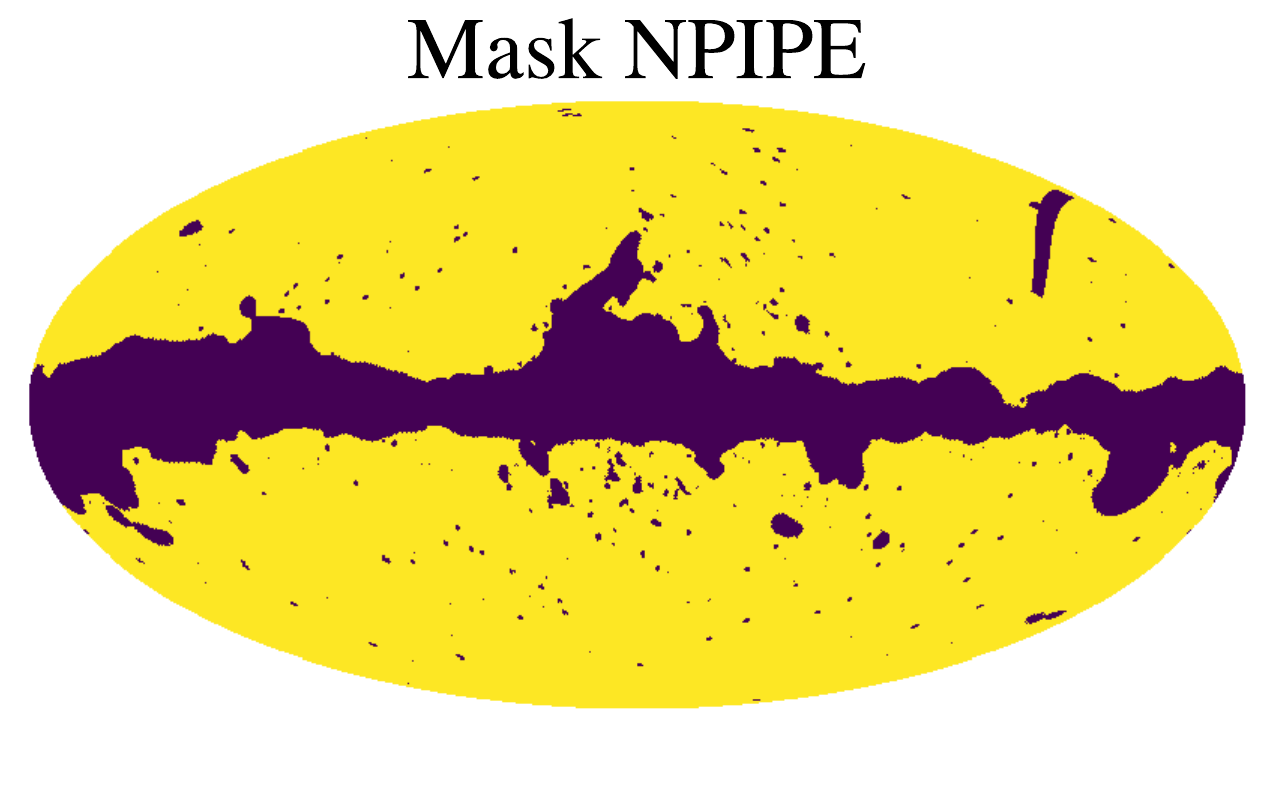}
    \includegraphics[width=.32\linewidth]{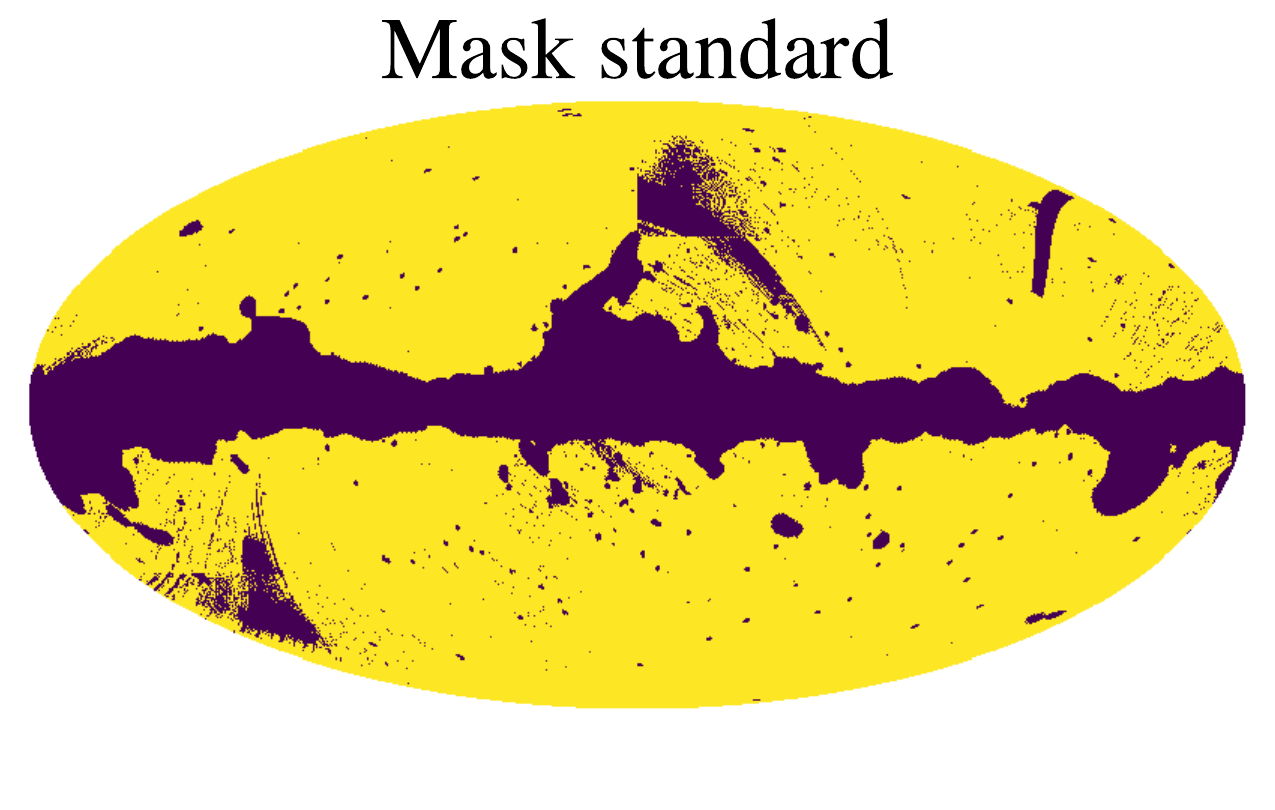} \\
    \includegraphics[width=.32\linewidth]{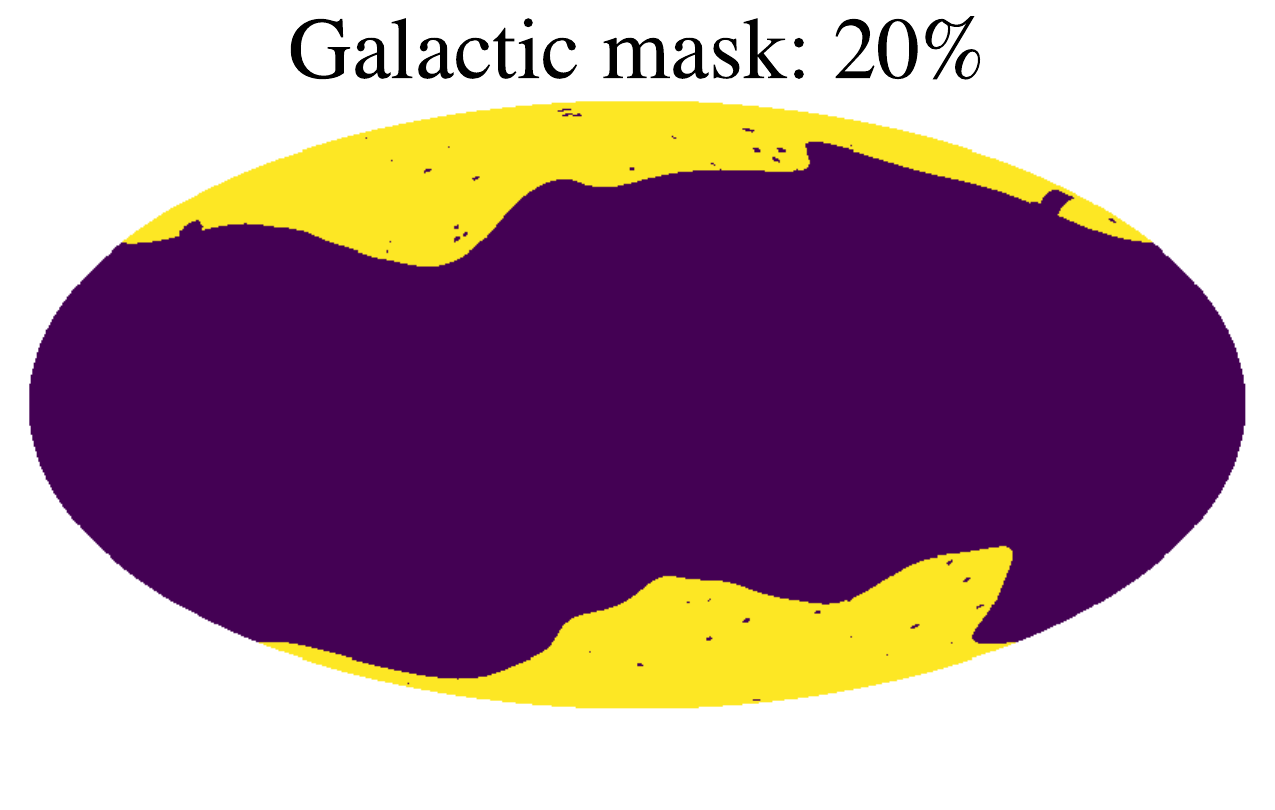} 
    \includegraphics[width=.32\linewidth]{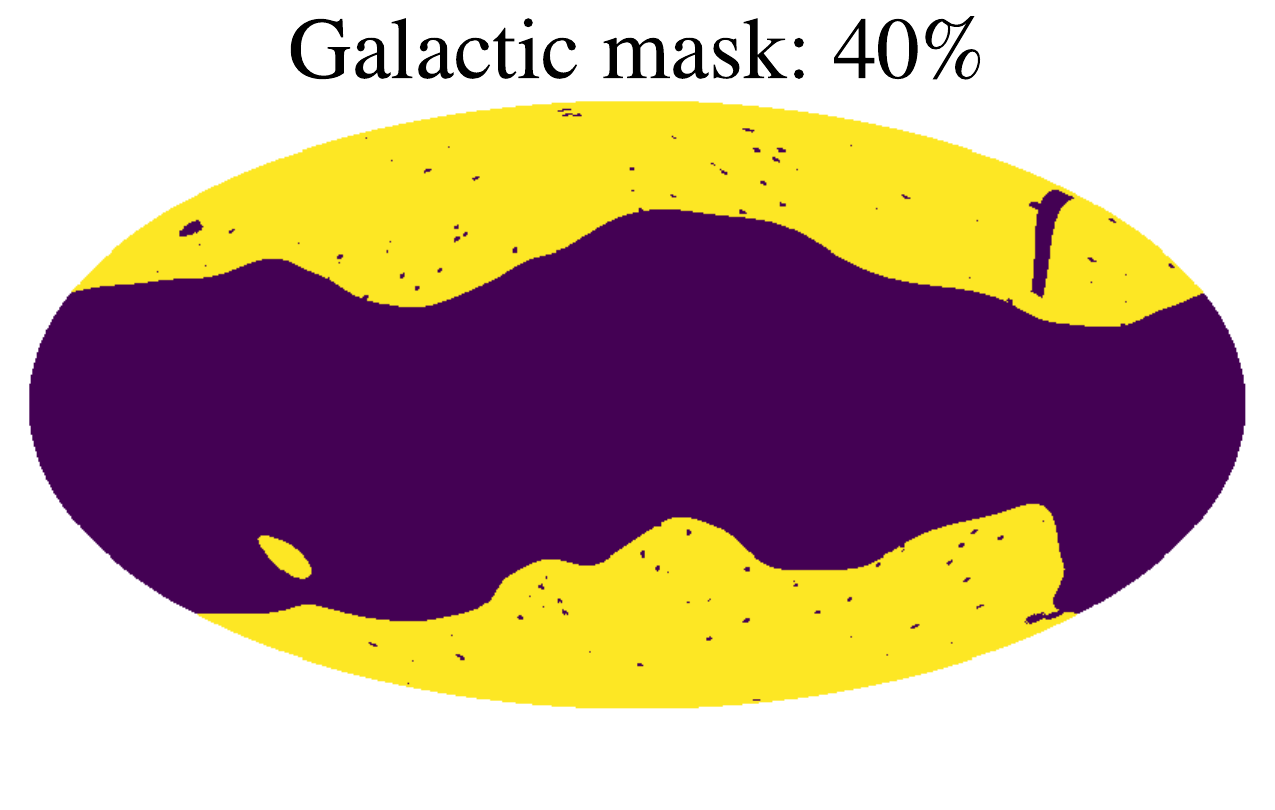} 
    \includegraphics[width=.32\linewidth]{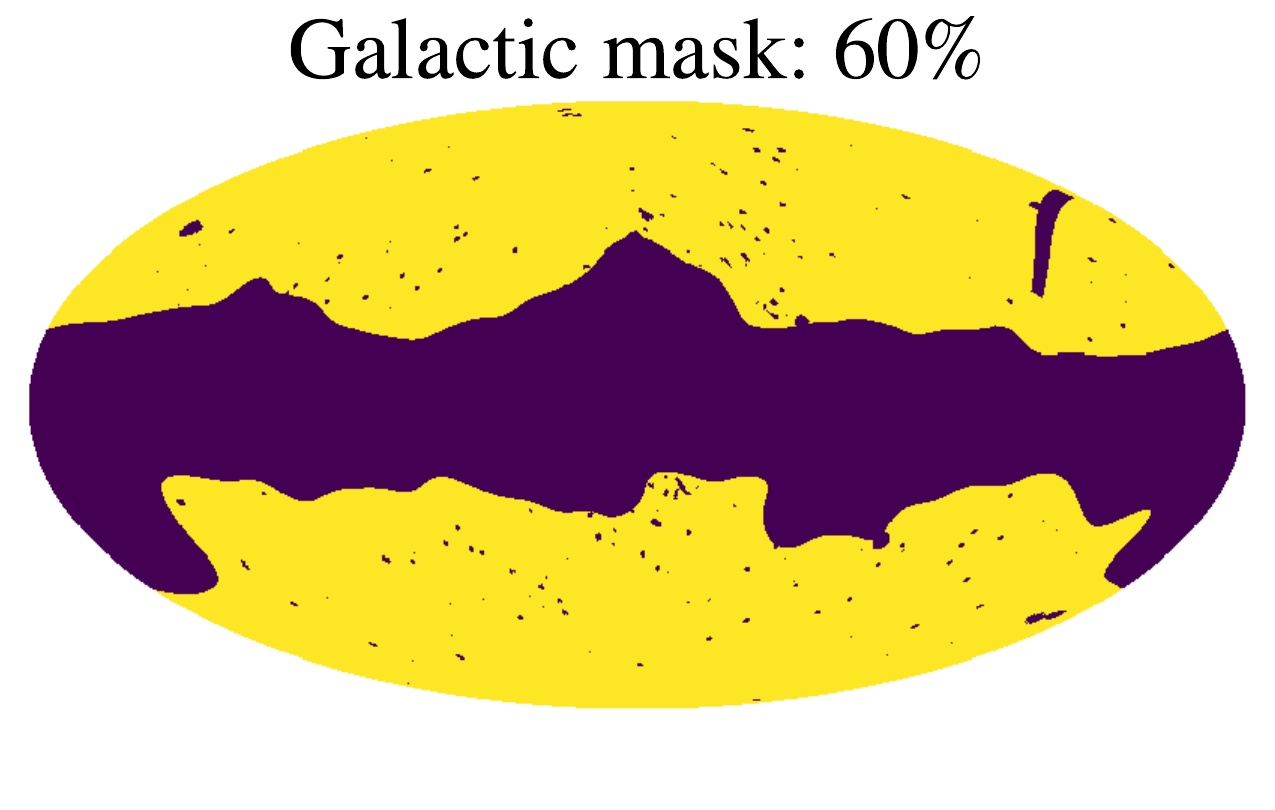}  
\caption{\label{fig:pol_mask} Different masks used in this work. From left to right: \planck\,NPIPE standard mask which retains $78\%$ of the sky; \planck\,PR3 standard mask which retains $75\%$ of the sky; NPIPE standard plus galactic mask with $20\%$ of sky coverage; NPIPE standard plus galactic mask with $40\%$ sky coverage; NPIPE standard plus galactic mask with $60\%$ sky coverage.}
\end{figure}
The CMB maps employed for the main analysis of \planck\,NPIPE data products are masked with the \planck\,fiducial analysis mask corresponding to a sky fraction of $f_{sky}=78\%$ (first mask in the first row of figure \ref{fig:pol_mask}), while the mask used for \planck\,PR3 data products is characterized by a sky fraction of $f_{sky}=75\%$ (second mask in the first row of figure \ref{fig:pol_mask}). As presented in section \ref{subsec:consistency checks}, we also test for different sky coverages, meaning for different masks applied to CMB data and simulations. Figure \ref{fig:pol_mask} shows the masks used in this work.

\section{Results} \label{sec:Results}
We now present the application of our pipeline to \planck\,\textit{full mission} data and to different choices of data splits. Our primary findings have been obtained from \planck\,NPIPE and PR3 data products, cleaned with the component separation method \commander. Furthermore, we present a consistency check comparing the CB power spectra obtained from \textit{full mission} data products cleaned with the other component separation methods, \textit{i.e.} \sevem, \smica\,and \nilc\,for \planck\,PR3 and only \sevem\, for \planck\,NPIPE. Subsequently, with the obtained $\alpha_{LM}$ estimates, we discuss the procedure and present the results to end up with the map of the CB field and its cross-correlations with the CMB temperature and polarization fields.\par
From now on, we use \textit{[A]}, \textit{[B]} superscripts to indicate whether a quantity has been evaluated from split A or B of \planck\,NPIPE data splits, respectively, and \textit{[1]} and \textit{[2]} superscripts to distinguish among the \textit{half-mission 1} or \textit{half-mission 2} of \planck\,PR3. If no superscript is present, it means that the quantity has been evaluated from \textit{full mission} data.\par
All the following results that are presented have been obtained using CMB polarization maps at the full resolution of \Nside = 2048. Furthermore, unless otherwise stated, we exclude from our analysis the first 50 CMB multipoles ($\ell_{min}^{CMB}=50$) and the maximum multipole included in the analysis is $\ell_{max}^{CMB}=2000$. The CB power spectrum has been evaluated from $L_{min}^{CB}=0$ up to $L_{max}^{CB}=1500$.\par
The main steps of the analysis are the following:
\begin{itemize}
    \item evaluate the first estimates of the spherical harmonic coefficients of the CB field (eq. \eqref{alpha_est}) from all the CMB maps, \textit{i.e.} both data and simulations;
    \item compute the mean field bias (eq. \eqref{mf bias}) averaging the $\overline{\alpha}_{LM}$ coefficients estimated from simulations only;
    \item subtract the mean field bias from all the $\overline{\alpha}_{LM}$ estimates, both the ones from data and that from simulations;
    \item calculate the $C_L^{\hat{\alpha}\hat{\alpha}}$ (eq.\eqref{cl_aa_map}) and the $C_L^{bias,iso}$ (eq.\eqref{bias_iso}) for both data and simulations of \planck\,data products;
    \item split the $C_L^{\hat{\alpha}\hat{\alpha}}$ and the $C_L^{bias,iso}$ evaluated from simulations only, into two equally sized sets:
    \begin{itemize}
        \item \underline{Set A}, used for the evaluation of the Monte Carlo bias term (eq.\eqref{bias_MC});
        \item \underline{Set B}, used to obtain a final set of fully de-biased simulations. In particular, for each simulation of this set, we evaluate the unbiased CB power spectrum, $\hat{C}_L^{\alpha\alpha}$, subtracting the isotropic bias term, $C_L^{bias,iso}$ evaluated from the same set, and the Monte Carlo bias term, $C_L^{bias,MC}$, calculated from \underline{set A};
    \end{itemize}
        \item subtract the isotropic bias term evaluated from data, $C_L^{bias,iso}$, and the Monte Carlo bias term evaluated from \underline{set A}, $C_L^{bias,MC}$, from the biased $\alpha\alpha$ power spectrum calculated on \planck\,data, in order to obtain the estimated CB power spectrum, $\hat{C}_L^{\alpha\alpha}$ (eq. \eqref{PS_est}).
\end{itemize}

\subsection{Full mission}
We present the application of our pipeline in case of a CB power spectrum obtained combining together $\alpha_{LM}$ coefficients estimated from \planck\,NPIPE \textit{full mission} data.\par
The equation for the $\alpha\alpha$ power spectrum before the de-bias is the same of equation \eqref{cl_aa_map}, where $f_{sky}$ is the sky fraction of the mask (first mask in the first row of figure \ref{fig:pol_mask}) applied to CMB polarization maps, both data and simulations, and corresponding to $f_{sky} = 0.78$.\par
\begin{figure}[ht]
\centering
\includegraphics[width=.7\textwidth]{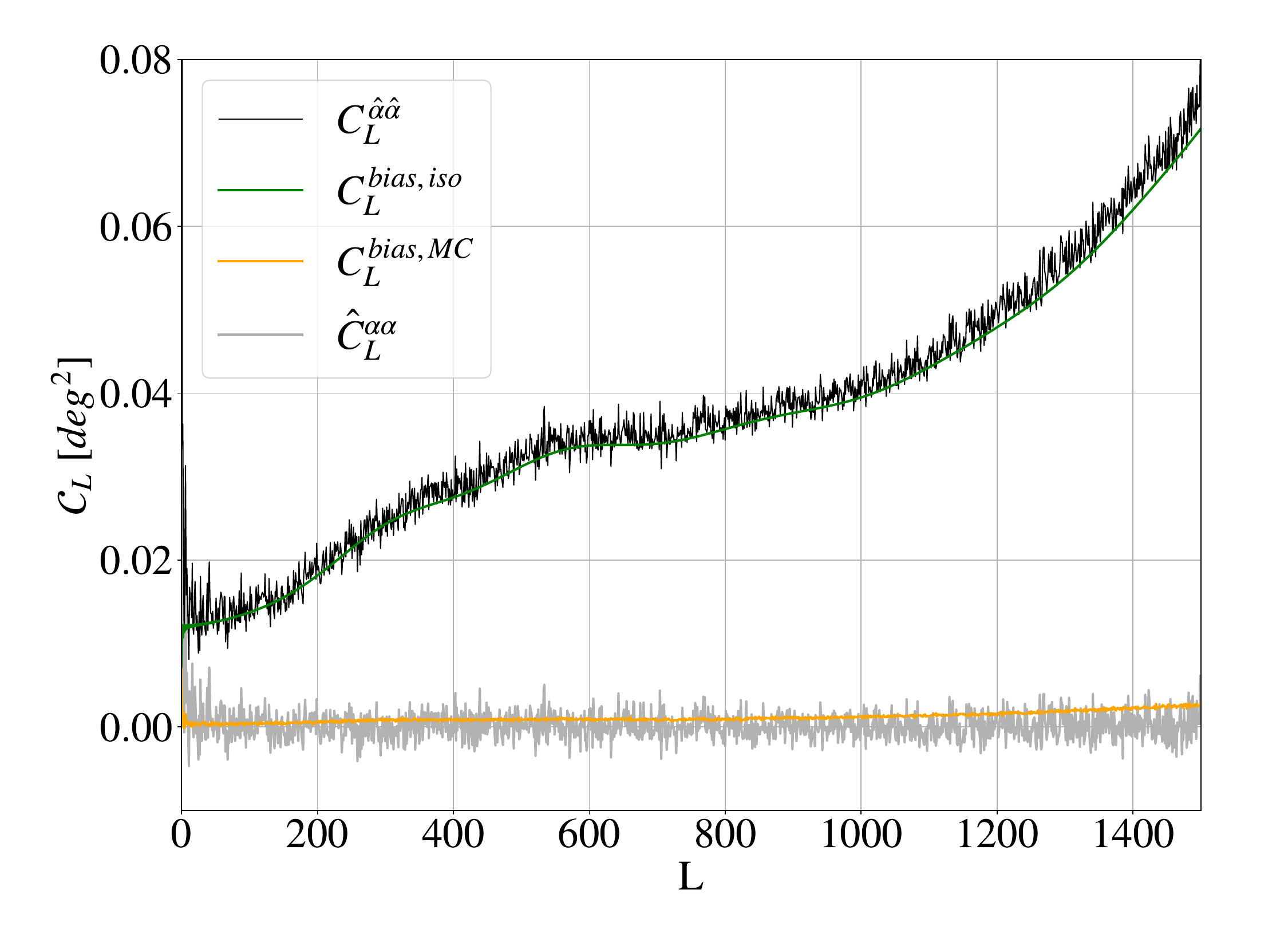}
\caption{\label{fig:auto bias fm}Power spectra evaluated from \textit{full mission} NPIPE data and simulations: $\alpha\alpha$ power spectrum before the de-biasing procedure, evaluated from \planck\,data (\textit{black curve}); isotropic bias term evaluated from \planck\,data (\textit{green curve}); Monte Carlo bias term evaluated from the simulation \underline{set A} (\textit{orange curve}); $\alpha\alpha$ power spectrum after the de-biasing procedure (\textit{gray curve}).}
\end{figure}
\begin{figure}[ht]
\centering
\includegraphics[width=.7\textwidth]{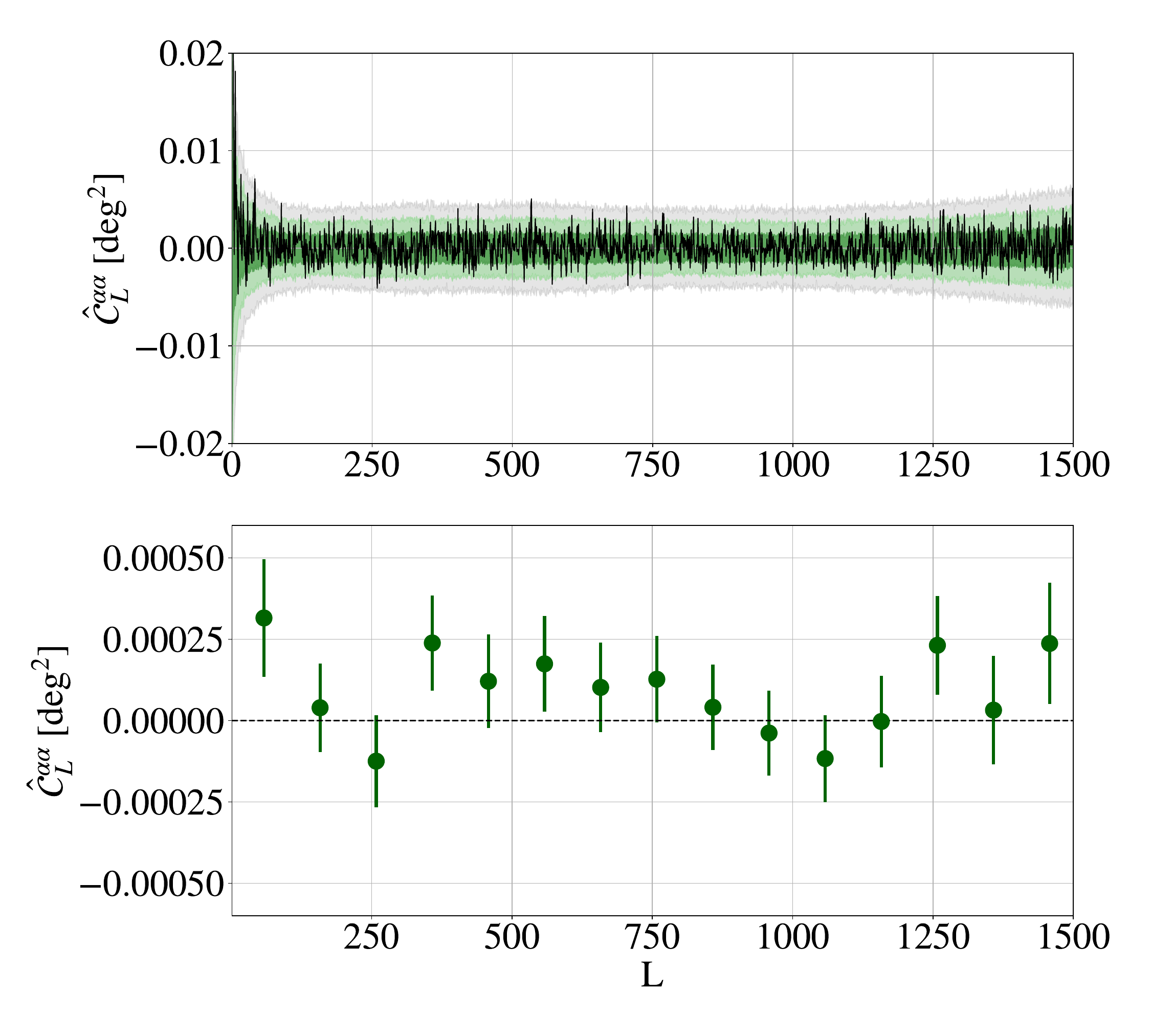}
\caption{\label{fig:auto debias fm}Results evaluated from \textit{full mission} NPIPE data. \underline{Upper panel} De-biased $\alpha\alpha$ power spectrum (\textit{black line}), with the $1\sigma$ (\textit{dark green area}), $2\sigma$ (\textit{light green area}) and $3\sigma$ (\textit{gray area}) confidence levels. \underline{Lower panel} De-biased $\alpha\alpha$ power spectrum after binning with 100 multipoles per bin and excluding the first 8 multipoles.}
\end{figure}

In figure \ref{fig:auto bias fm} we plot the power spectra of the different terms. The black and green curves represent the biased $\alpha\alpha$ power spectrum and the \textit{isotropic bias term}, respectively, evaluated using the $\alpha_{LM}$ estimates from \planck\,NPIPE data, while the orange curve represents the \textit{Monte Carlo bias term}, computed by averaging the de-biased $\alpha\alpha$ power spectra evaluated over the first 200 \textit{CMB+noise} simulations, that is, evaluated from the simulation \underline{set A}. Even though we show the estimate of the $\alpha\alpha$ power spectrum in figure \ref{fig:auto debias fm}, we display it with the gray curve also in figure \ref{fig:auto bias fm}.\par
The black curve in the upper panel of figure \ref{fig:auto debias fm} is the $\alpha\alpha$ power spectrum evaluated for each multipole after the de-biasing procedure (eq. \eqref{PS_est}). The shaded areas represent the $1\sigma$, $2\sigma$ and $3\sigma$ confidence intervals, obtained after the computation of the variance of the fully de-biased $\alpha\alpha$ power spectra evaluated over the simulation \underline{set B}. In order to better visualize the results, in the lower panel of figure \ref{fig:auto debias fm} we also plot the de-biased CB power spectrum after binning with 100 multipoles per bin, excluding the first 8 multipoles; the error bars are at $1\sigma$.\par
In figure \ref{fig:auto bias fm} the rotation signal induced by CB before the de-biasing procedure, \textit{i.e.} $C_L^{\hat{\alpha}\hat{\alpha}}$, is different from zero. We can understand the reason of this strong deviation from zero since the single estimate for the $\alpha_{LM}$ coefficients goes as:
\begin{equation}
    \hat{\alpha}_{LM} \propto a_{\ell m}^{E, map}a_{\ell'm'}^{B,map,*} ,
\end{equation}
and when combining two estimates to obtain the power spectrum before the de-bias, we end up with:
\begin{equation}\label{claa_behaviour}
    C_L^{\hat{\alpha}\hat{\alpha}} \propto (a_{\ell_1 m_1}^{E, map}a_{\ell_2m_2}^{B,map,*})(a_{\ell_3 m_3}^{E, map,*}a_{\ell_4m_4}^{B,map}) . 
\end{equation}
The only non negligible contributions come from:
\begin{align} 
    a_{\ell_1 m_1}^{E,map}a_{\ell_3 m_3}^{E,map,*} \sim C_{\ell_1}^{EE,map} ,\\
    a_{\ell_2 m_2}^{B,map}a_{\ell_4 m_4}^{B,map,*} \sim C_{\ell_2}^{BB,map}.   
\end{align}
The observed EE and BB power spectra encode for two contributions; the cosmological signal, \textit{i.e.} $C_\ell^{EE}$ and $C_\ell^{BB}$, and the noise contribution, \textit{i.e} $N_\ell^{EE}$ and $N_\ell^{BB}$, so that:
\begin{align} 
    C_\ell^{EE,map} = C_\ell^{EE} + N_\ell^{EE}, \\
    C_\ell^{BB,map} = C_\ell^{BB} + N_\ell^{BB}.   
\end{align}
In case of $a_{\ell m}^{X,map}$ observed from the same data set, the noise auto-correlates. For the \Planck\ sensitivity the dominant contribution to the observed EE and BB power spectra comes from the noise itself, and the bias in Fig.~\ref{fig:auto bias fm} is dominated by the auto-correlation of the noise. Nevertheless, it is worth stressing that even in a signal dominated regime the de-bias procedure is necessary for removing the signal auto-correlation.\par
The CB power spectrum after the de-biasing procedure obtained applying our pipeline to \textit{full mission} \planck\,NPIPE data is compatible with zero with a Probability To Exceed (PTE) of $84.35\%$.\par
\begin{figure}[ht]
\centering
\includegraphics[width=.49\textwidth]{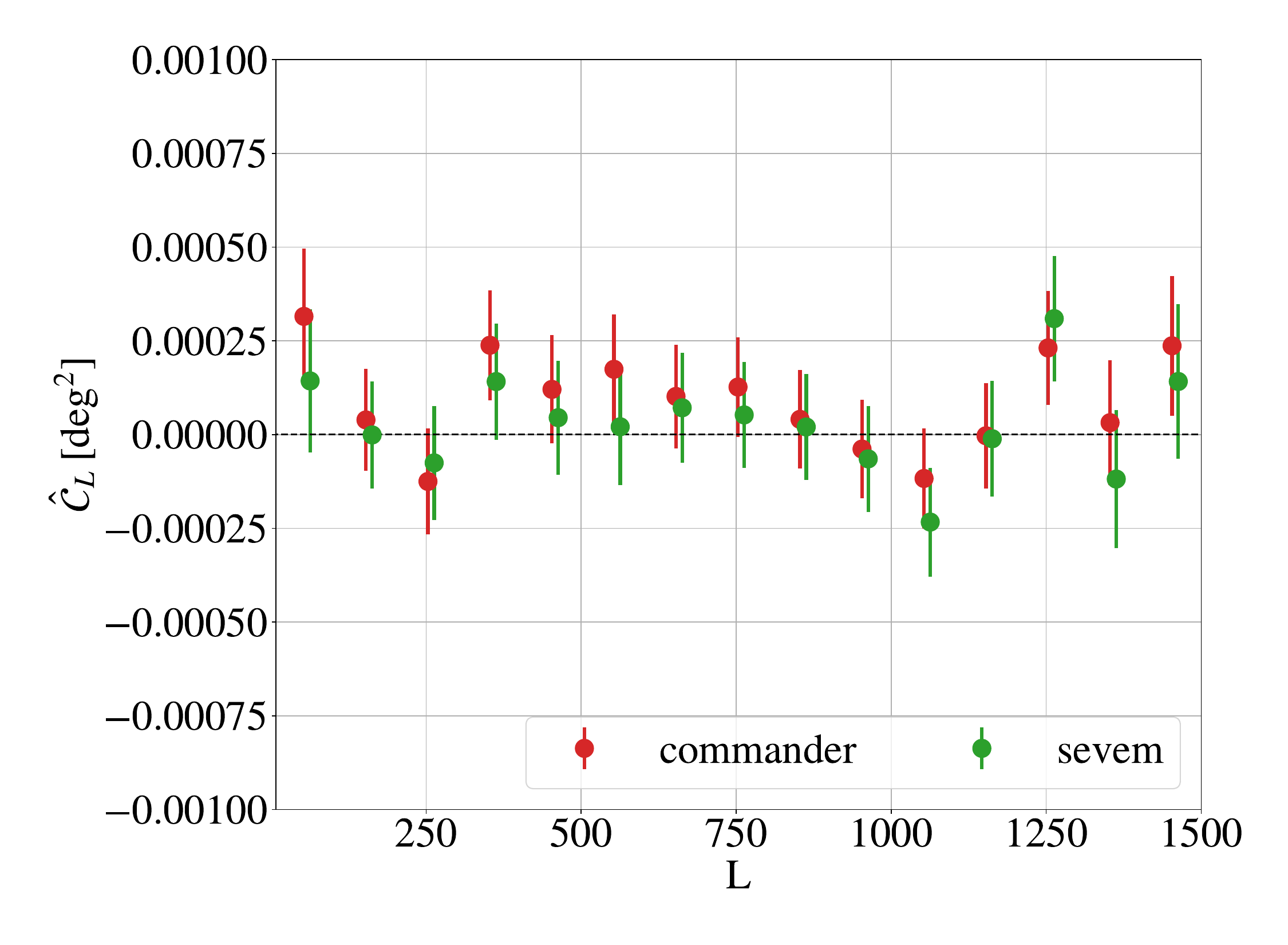} 
\includegraphics[width=.49\textwidth]{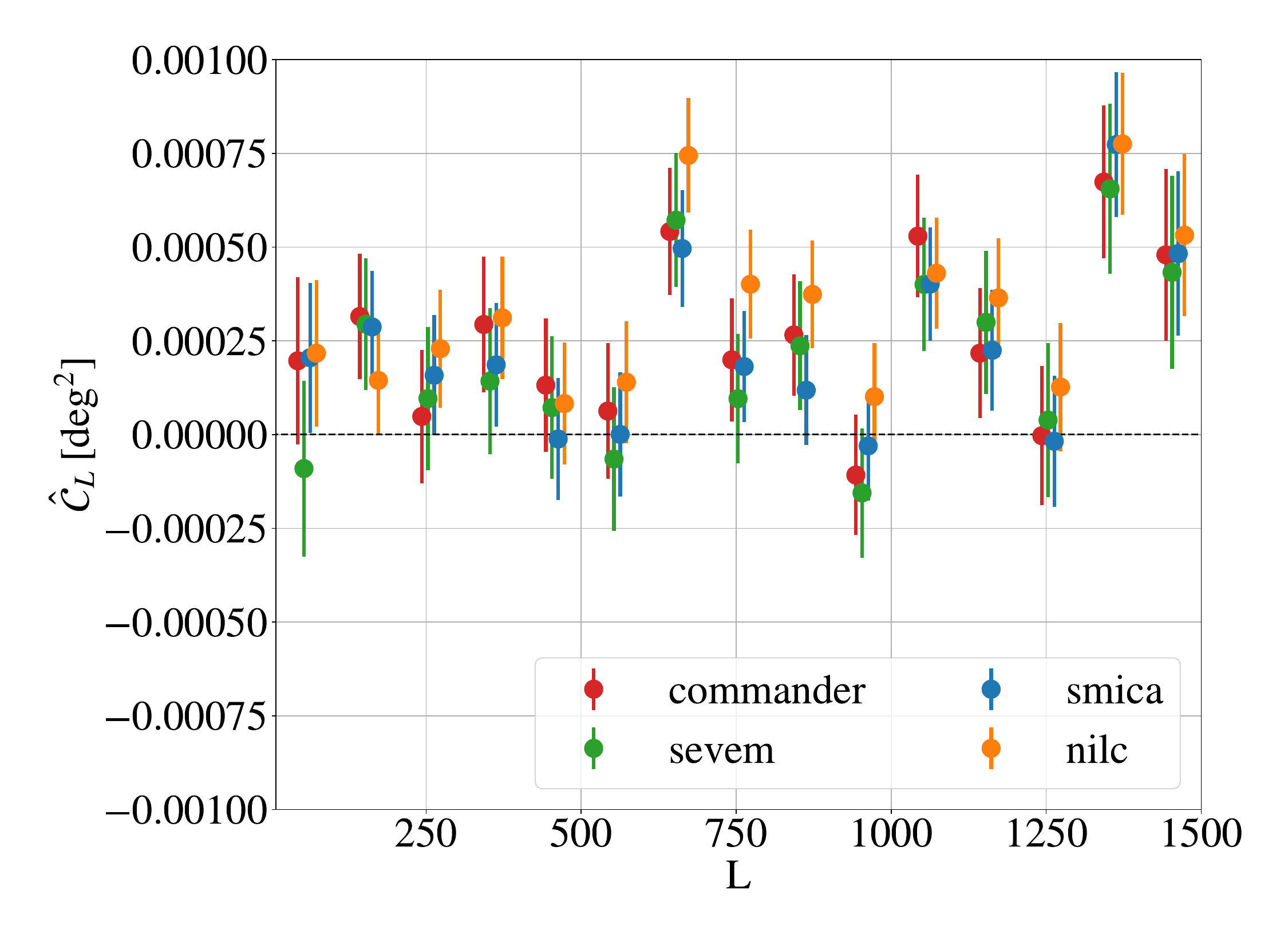}    
\caption{\label{fig:comp sep PR3 NPIPE}\underline{Left panel} De-biased $\alpha\alpha$ power spectrum after binning with 100 multipoles per bin and excluding the first 8 multipoles using \planck\,NPIPE \textit{full mission} data. We compare the different component separation methods; \commander\,in red and \sevem\,in green. \underline{Right panel} Same as the left panel, but for \planck\,PR3 data and for the four component separation methods available for PR3; in blue we show the power spectrum obtained using \smica\, component separation method, in orange \nilc, in green \sevem\, and in red \commander.}
\end{figure}

In the left and right panels of figure \ref{fig:comp sep PR3 NPIPE} we show the de-biased $\alpha\alpha$ power spectrum evaluated from \textit{full mission} \planck\,NPIPE and PR3 data products respectively, for the different component separation methods. In the left panel of figure \ref{fig:comp sep PR3 NPIPE} we show the de-biased power spectrum after binning with 100 multipoles per bin and excluding the first 8 multipoles, evaluated from \planck\,NPIPE data products, for the two component separation methods available, \textit{i.e.}, for \commander, in red, and for \sevem, in green. The estimated CB power spectrum is consistent among the different component separation methods. In the right panel of figure \ref{fig:comp sep PR3 NPIPE} we show the same de-biased $\alpha\alpha$ power spectrum evaluated from \planck\,PR3 data products for the four component separation methods, meaning \smica\,(blue), \nilc\,(orange), \sevem\,(green) and \commander\,(red). The CB power spectrum estimated from \planck\,PR3 \commander\,is compatible with zero with a PTE of $6.61\%$. This low PTE reflects the behaviour observed in the data at large multipoles (see right panel of figure \ref{fig:comp sep PR3 NPIPE}), where there appears to be an excess of power. We speculate that this could be addressed to a mismatch between the noise in the CMB simulations and the one of data of \planck\,PR3 data products.\par
\begin{figure}[ht]
\centering
\includegraphics[width=.49\textwidth]{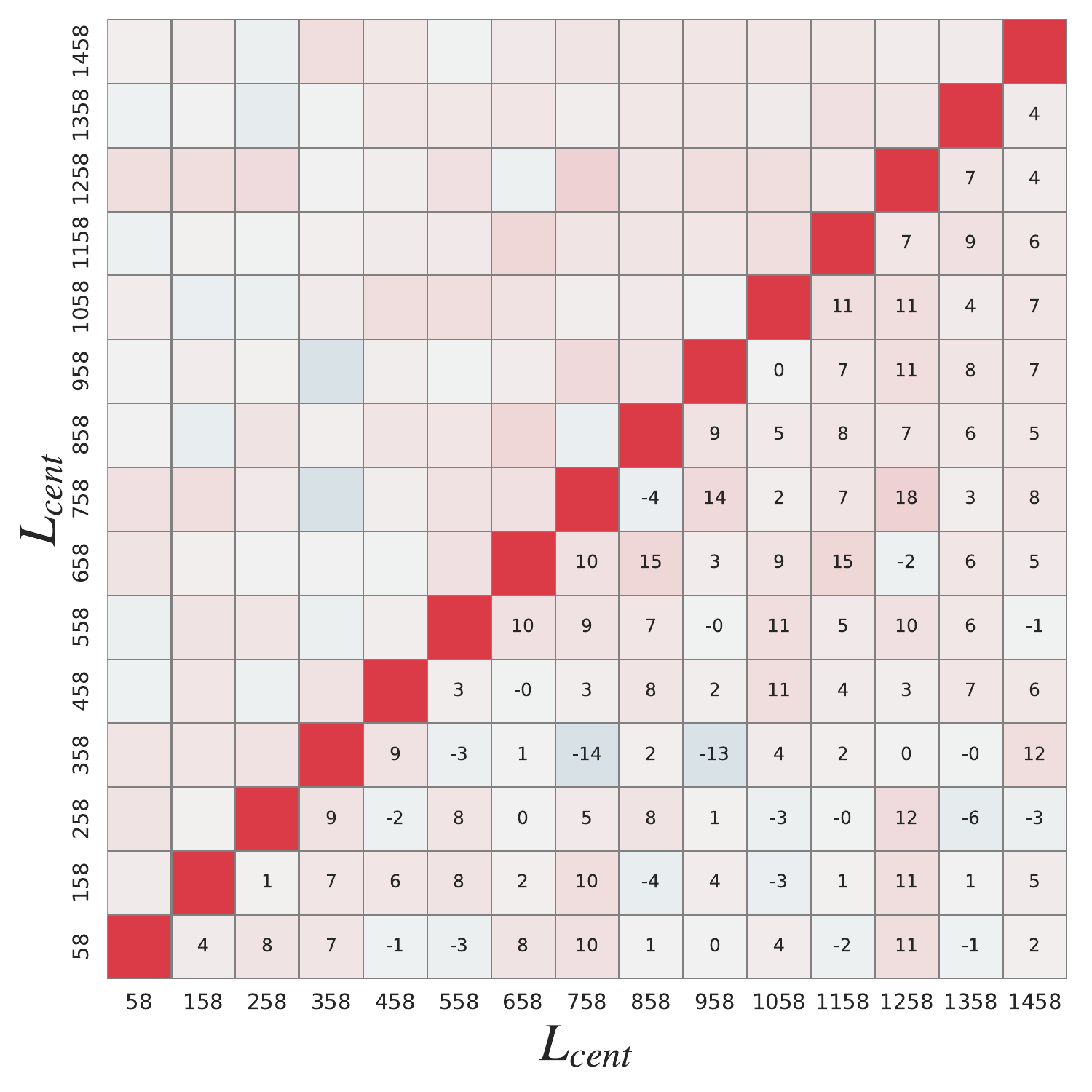}
\includegraphics[width=.49\textwidth]{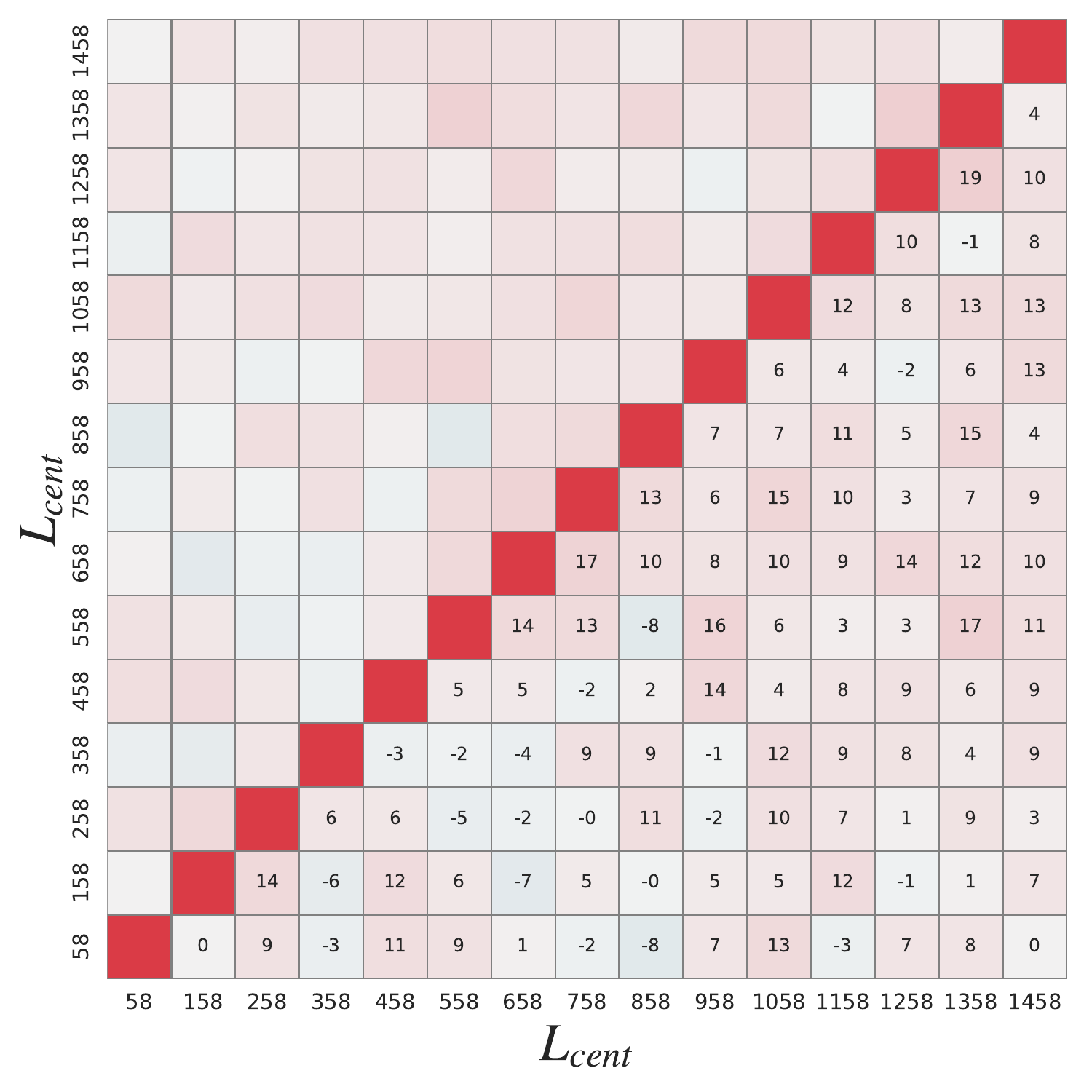}
\caption{\label{fig:corr matrix fm}\underline{Left panel} Correlation matrix evaluated from the 400 CB power spectra estimated from the simulations of \planck\,NPIPE data products. \underline{Right panel} Correlation matrix evaluated from the 300 CB power spectra estimated from the simulations of \planck\,PR3 data products. All simulations are binned with 100 multipoles per bin. The correlation coefficients are expressed in terms of percentage.}
\end{figure}

In the left and right panels of figure \ref{fig:corr matrix fm} we also show the correlation matrix evaluated from the $\alpha\alpha$ power spectra estimated from the 400 CMB+noise simulations of \planck\,NPIPE and from the 300 CMB+noise simulations of \planck\,PR3, respectively, after binning with 100 multipoles per bin. The multipole $L_{cent}$ reported in the axes of the matrix is the center of the multipole bin. The correlation coefficients reported in figure \ref{fig:corr matrix fm} are expressed in terms of percentage.\par

\subsection{Data splits}
We summarize the results of the application of our pipeline to different combinations of data splits. In particular, we show the CB power spectrum obtained:
\begin{itemize}
    \item auto-correlating $\alpha_{LM}$ coefficients estimated from data split A (\textit{half mission 1}) of \planck\, NPIPE (PR3) data products;
    \item auto-correlating $\alpha_{LM}$ coefficients estimated from data split B (\textit{half mission 2}) of \planck\, NPIPE (PR3) data products;
    \item cross-correlating $\alpha_{LM}$ coefficients evaluated from data split A (\textit{half mission 1}) and data split B (\textit{half mission 2}) of \planck\,NPIPE (PR3) data products.
\end{itemize}
A word of caution concerning the case of the cross-correlation (since both auto-correlations follow exactly the same pipeline of the \textit{full mission} case). The main differences with respect to the previously described case are the following:
\begin{itemize}
    \item the calculation of the $\alpha\alpha$ power spectrum before the debias (eq. \eqref{cl_aa_map}), as well as the one of the isotropic bias term (eq. \eqref{bias_iso}), involves estimates of the $\alpha_{LM}$ coefficients coming from the two data splits (or from the two half-missions, if we work with \planck\,PR3 data products);
    \item all quantities involving window functions and noise curves, \textit{i.e.} $F_{\ell\ell'}^{L,EB}$, $F_{\ell\ell'}^{L,BE}$ and the analytic power spectra, $C_\ell^{XX,map}$, must be evaluated separately for each data split (or each half-mission).
\end{itemize}
Thus, the biased power spectrum of equation \eqref{cl_aa_map} is now calculated as:
\begin{equation}\label{alpha_est_cc}
    C_L^{\hat{\alpha}\hat{\alpha}} = \dfrac{1}{f_{sky}}\dfrac{1}{2L+1}\displaystyle\sum_M \hat{\alpha}_{LM}^{[A]}\hat{\alpha}_{LM}^{[B],*} .
\end{equation}
And, since we are combining together estimates coming from two different data sets, the normalization of the $\hat{\alpha}_{LM}$ coefficients is different depending on whether we are dealing with data split A (\textit{half-mission 1}) or data split B (\textit{half-mission 2}) estimates. The distinction follows the same notation, \textit{i.e.} we indicate the inverse variance from the data split A (\textit{half-mission 1}) as $(\sigma_L^{-2})^{[A]}$ ($(\sigma_L^{-2})^{[1]}$) and as $(\sigma_L^{-2})^{[B]}$ ($(\sigma_L^{-2})^{[2]}$) the one from the data split B (\textit{half-mission 2}).\par
The same modification applies to the isotropic bias term \eqref{bias_iso}, which now reads as:
\begin{equation}\label{bias_iso_cc}
    C_L^{bias,iso} = <\hat{\alpha}_{LM}^{[A]}\hat{\alpha}_{LM}^{[B],*}> .
\end{equation}
\begin{figure}[ht]
\centering
\includegraphics[width=.49\textwidth]{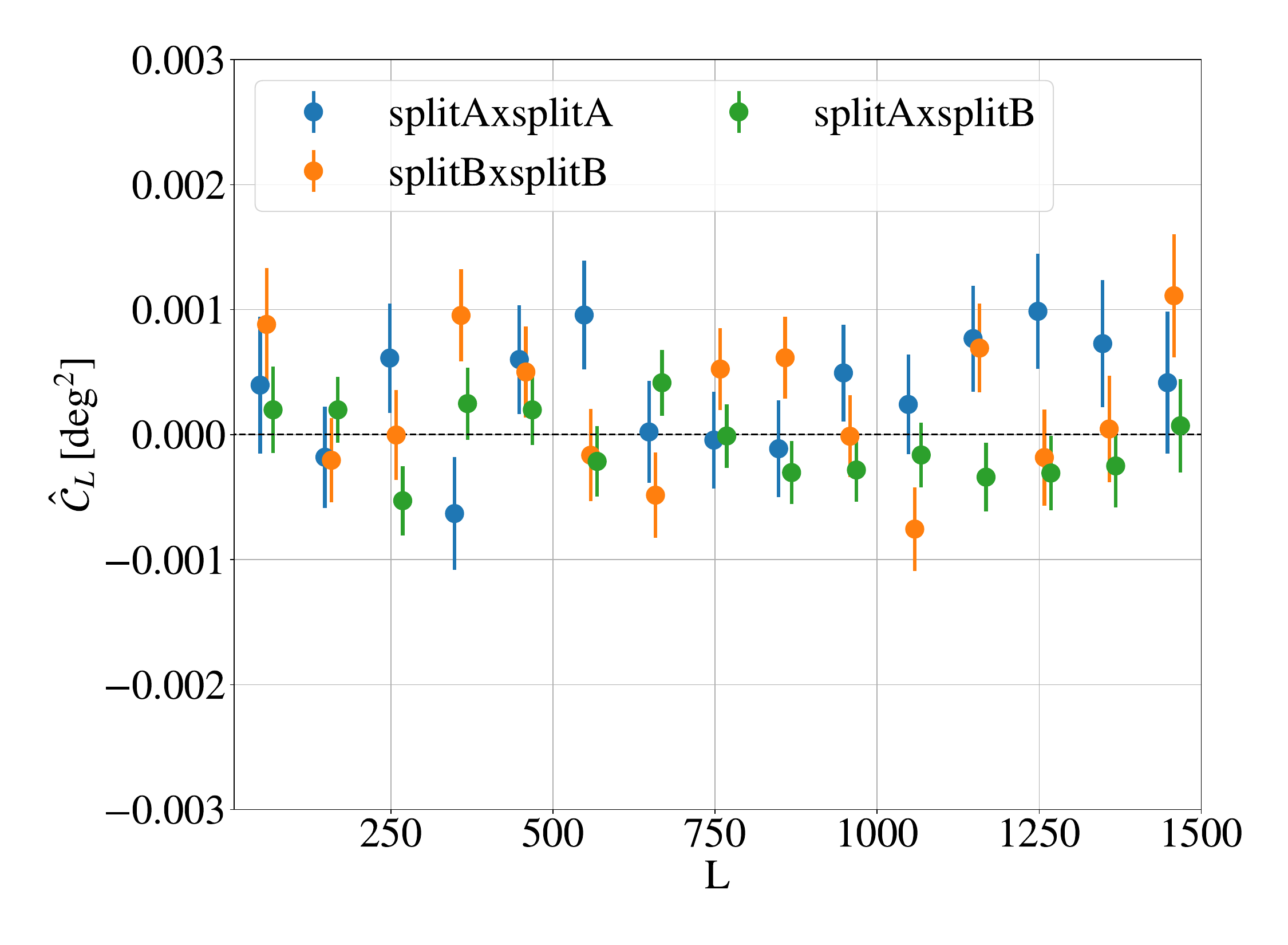}
\includegraphics[width=.49\textwidth]{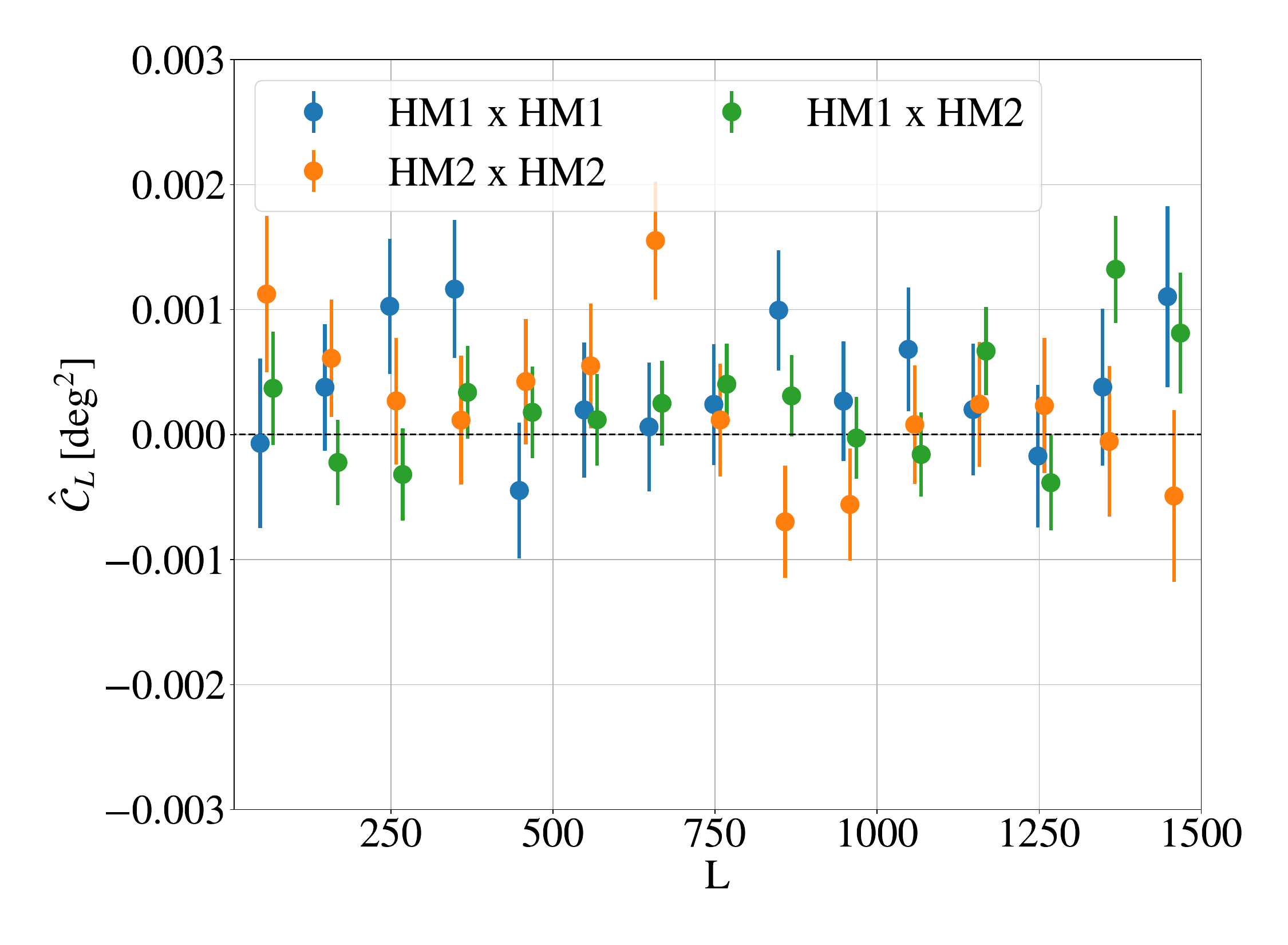}
\caption{\label{fig:PR3 and PR4 spectra}\underline{Left panel} De-biased $\alpha\alpha$ power spectrum after binning with 100 multipoles per bin, excluding the first 8 multipoles from the analysis, obtained auto-correlating $\alpha_{LM}$ estimates from data split A (blue), data split B (orange) and cross-correlating estimates from the two data splits (green) of \planck\,NPIPE data products. \underline{Right panel} Same as the left panel, for the auto-correlation of $\alpha_{LM}$ estimates from \textit{half mission 1} (blue), \textit{half mission 2} (orange) and for the cross-correlation from the two half missions (green) of \planck\,PR3 data products.}
\end{figure}
In the left panel of figure \ref{fig:PR3 and PR4 spectra} we show the de-biased CB power spectrum after binning with 100 multipoles per bin, excluding the first 8 multipoles, for the different combinations of data splits of \planck\,NPIPE; the error bars are at $1\sigma$. More precisely, the presented results show the $\alpha\alpha$ power spectra obtained auto-correlating $\alpha_{LM}$ estimates from \textit{split A} (\textit{blue}), auto-correlating estimates from \textit{split B} (\textit{orange}) and cross-correlating $\alpha_{LM}$ estimates from both \textit{split A} and \textit{split B} (\textit{green}).\par
In the right panel of figure \ref{fig:PR3 and PR4 spectra} we show the application of the pipeline to the different choices of data splits of \planck\,PR3 data products; the auto-correlation from \textit{half mission 1} (\textit{blue}), the auto-correlation from \textit{half mission 2} (\textit{orange}) and the cross-correlation between \textit{half mission 1} and \textit{half mission 2} (\textit{green}).\par
The results presented in figure \ref{fig:PR3 and PR4 spectra} have been obtained using the \planck\,NPIPE and PR3 data products cleaned with the component separation method \commander. In this section we present the results for \commander\,only since our analysis shows consistency among the different component separation methods (see left and right panels of figure \ref{fig:comp sep PR3 NPIPE}). The CB power spectra estimated for the different data splits are compatible with zero with the PTE reported in table \ref{tab: PTEs}.\par
\begin{table}[ht]
\centering
\renewcommand{\arraystretch}{1.5} 
\begin{tabular}{| c | c || c | c |}
 \hline
 \multicolumn{2}{|c||}{NPIPE \commander} & \multicolumn{2}{c|}{PR3 \commander} \\[1ex] \hhline{|=|=||=|=|} 
 $\alpha^{[A]}\alpha^{[A]}$& 71.01\% & $\alpha^{[1]}\alpha^{[1]}$& 75.95\%\\[1ex]
 $\alpha^{[B]}\alpha^{[B]}$& 7.77\% & $\alpha^{[2]}\alpha^{[2]}$& 50.48\%\\[1ex]
 $\alpha^{[A]}\alpha^{[B]}$& 83.08\% & $\alpha^{[1]}\alpha^{[2]}$& 31.88\%\\[1ex]
 \hline
 $\alpha\alpha$& 84.35\% & $\alpha\alpha$& 6.61\%\\[1ex]
 \hline
\end{tabular}
\caption{Probability To Exceed for the CB power spectra estimated for the different data splits presented in figure \ref{fig:PR3 and PR4 spectra}. For completeness, in the last row we also report the PTEs for the CB power spectra estimated from \planck\,NPIPE and \planck\,PR3 \textit{full mission} data products.}
\label{tab: PTEs}
\end{table}
Regarding the lowest PTEs listed in table \ref{tab: PTEs}, specifically those associated with $\alpha^{[2]}\alpha^{[2]}$ and $\alpha^{[B]}\alpha^{[B]}$, it is crucial to recognize that they correspond to distinct types of data splits. In the case of \planck\,PR3, the splits are time-based, whereas for \planck\,NPIPE we are working with detector-based data splits.\par

\subsection{Consistency checks}\label{subsec:consistency checks}
In the following, we consider the specific case of \textit{full mission} \planck\,NPIPE data products and we go through three consistency checks. In particular, we compare the de-biased $\alpha\alpha$ power spectrum:
\begin{itemize}
    \item for four different choices of the minimum CMB multipole included in the analysis, considering the cases where $\ell_{min}^{CMB} = 10$, $\ell_{min}^{CMB} = 30$, $\ell_{min}^{CMB} = 50$ and $\ell_{min}^{CMB} = 100$;
    \item for three different choices of the maximum CMB multipole included in the analysis, encoding for $\ell_{max}^{CMB} = 1500$, $\ell_{max}^{CMB} = 2000$ and $\ell_{max}^{CMB} = 2500$;
    \item for different masks applied to \planck\,NPIPE CMB polarization maps, corresponding to the sky-fractions of $f_{sky} = 78.0 \%, 59.2 \%, 39.7 \% , 20.3 \%$.
\end{itemize}
\begin{figure}[ht]
\centering
\includegraphics[width=.49\textwidth]{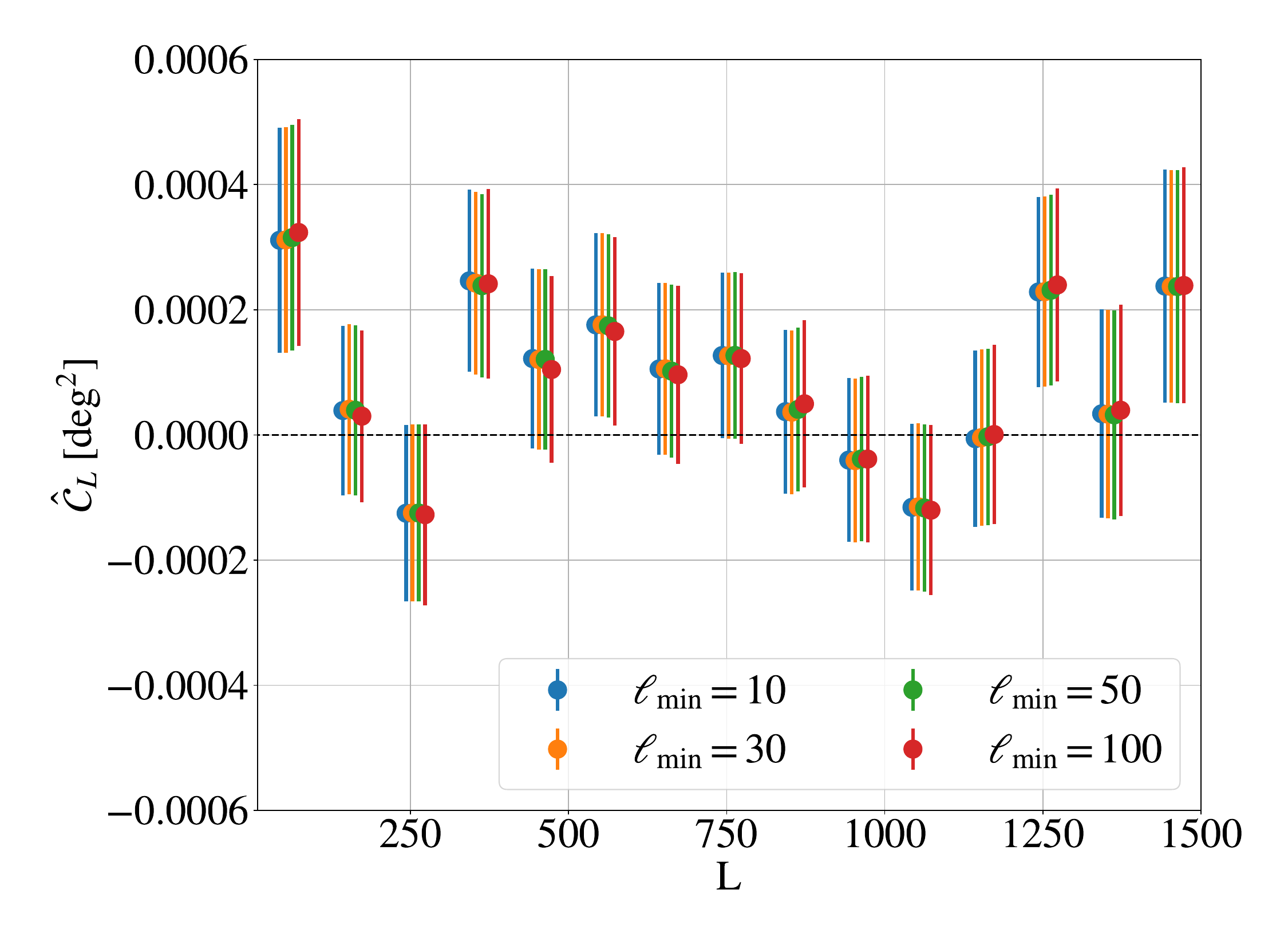}
\includegraphics[width=.49\textwidth]{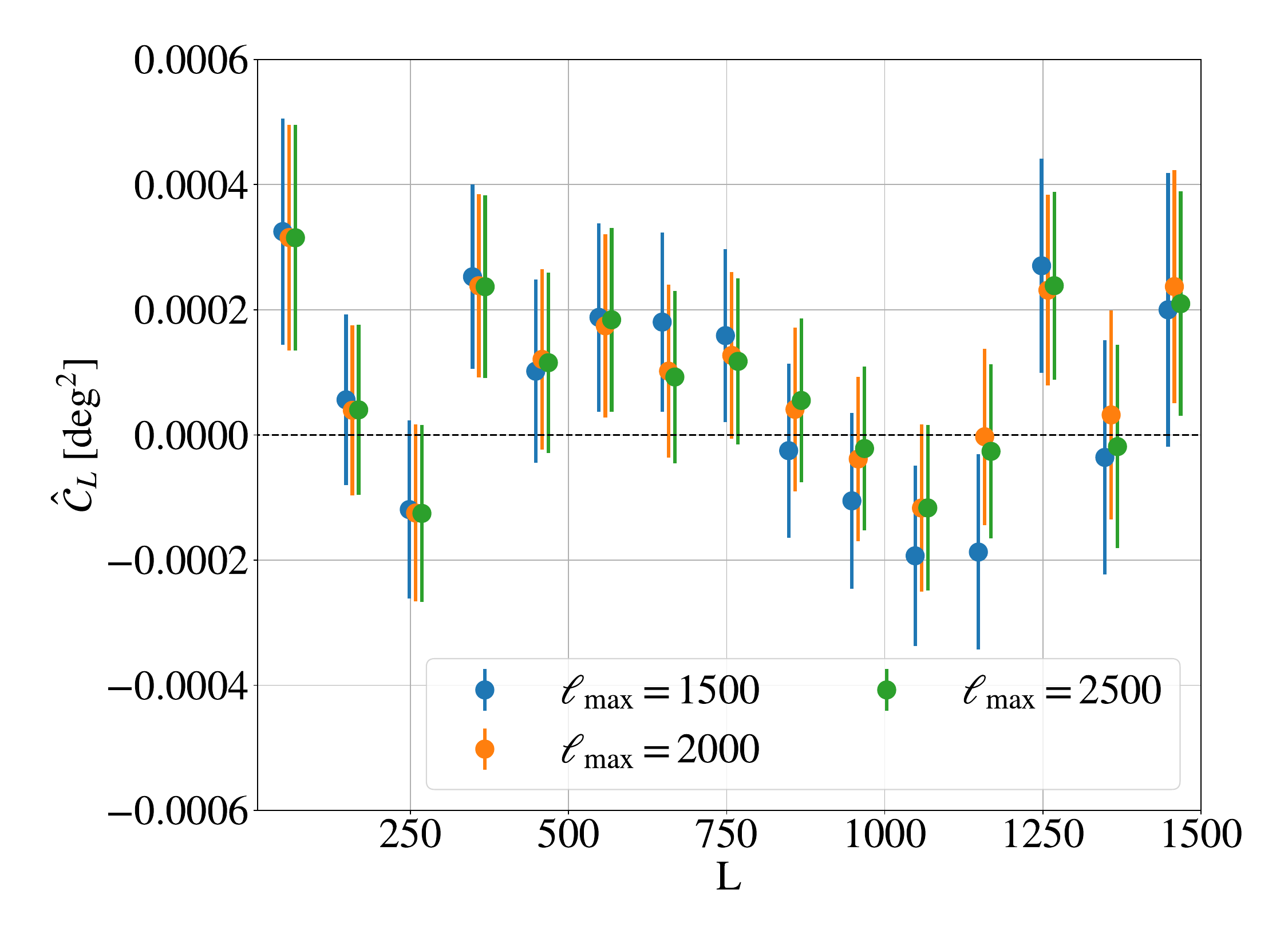}
\caption{\label{fig:consistency_check_ellCMB}CB power spectrum evaluated from \planck\,NPIPE \textit{full mission} data products, binning with 100 multipoles per bin and excluding the first 8 multipoles. \underline{Left panel} Power spectra for different values of $\ell_{min}^{CMB}$ included in the analysis and $\ell_{max}^{CMB} = 2000$ for all cases. We indicate in \textit{blue} the power spectrum obtained with $\ell_{min}^{CMB}=10$, in \textit{orange} the one with $\ell_{min}^{CMB}=30$, in \textit{green} the one with $\ell_{min}^{CMB}=50$ and in \textit{red} the one with $\ell_{min}^{CMB}=100$. \underline{Right panel} Power spectra for different values of $\ell_{max}^{CMB}$ included in the analysis and $\ell_{min}^{CMB} = 50$ for all cases. We indicate in \textit{blue}, \textit{orange} and \textit{green} the CB power spectra obtained including CMB multipoles up to $\ell_{max}^{CMB}=1500$, $\ell_{max}^{CMB}=2000$ and $\ell_{max}^{CMB}=2500$, respectively.}
\end{figure}
\begin{figure}[ht]
\centering
\includegraphics[width=.49\textwidth]{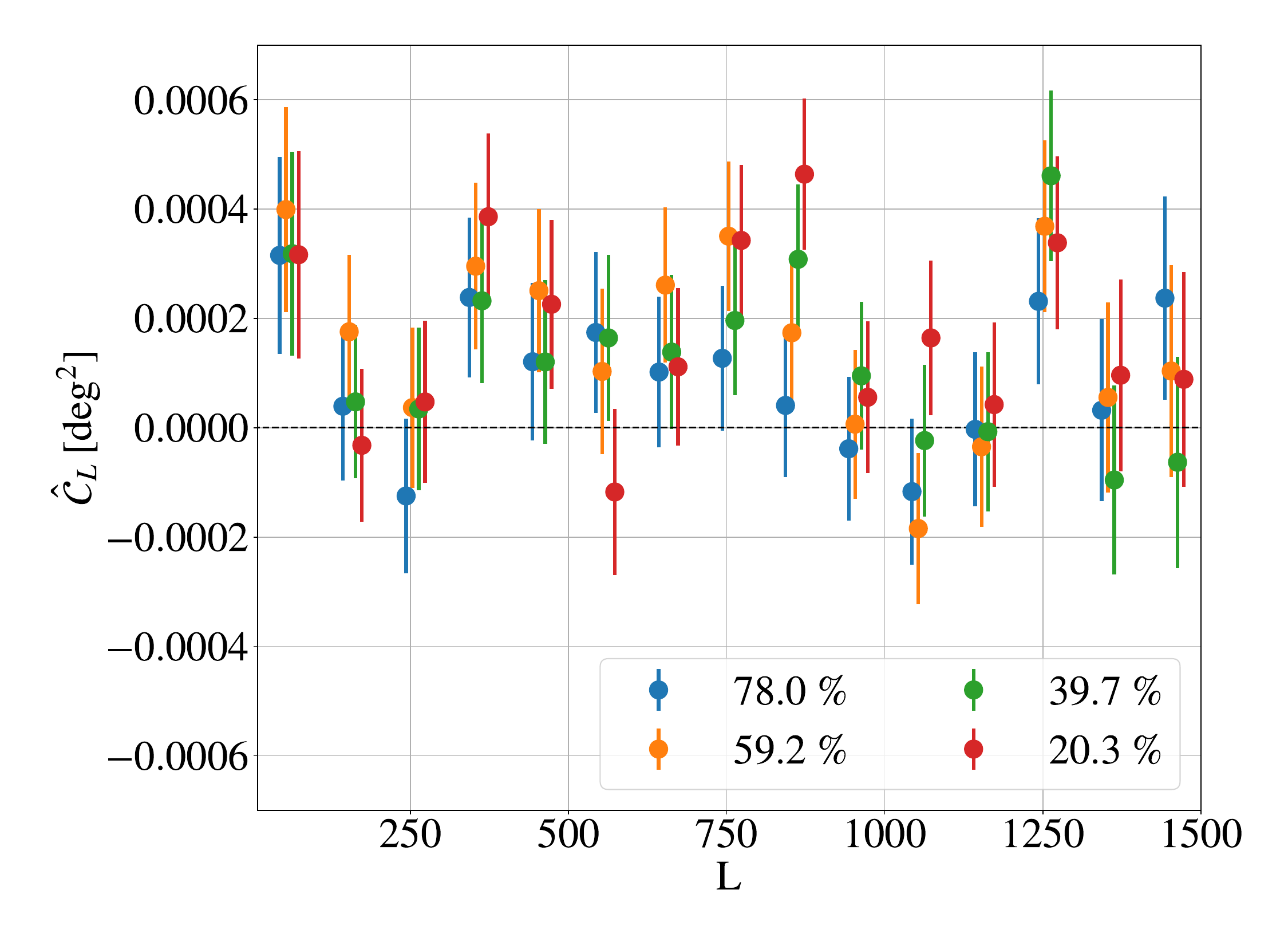}
\caption{\label{fig:consistency_check_mask}CB power spectrum evaluated from \planck\,NPIPE \textit{full mission} data products, binning over 100 multipoles per bin and excluding the first 8 multipoles, for different masks applied to \planck\,maps. In \textit{blue} we show the power spectrum obtained with the standard mask, corresponding to a $f_{sky}=78.0\%$; in \textit{orange}, \textit{green} and \textit{red} we show the power spectrum obtained applying masks corresponding to $f_{sky} = 59.2 \%, 39.7 \% , 20.3 \%$ respectively.}
\end{figure}
With reference to figure \ref{fig:consistency_check_ellCMB}, it is possible to conclude that the obtained results exhibit consistency across various choices of minimum and maximum CMB multipole included in the analysis.\par
The CB power spectrum results consistent also for the different masks applied to CMB data and simulations, as displayed in figure \ref{fig:consistency_check_mask}. It is worth noting that, despite the results showing strong consistency across different masks, the entire analysis normalized the estimator using its analytic variance instead of the true variance. As we observe a smaller portion of the sky, the difference between the analytic and the true variance becomes more pronounced. Specifically, we start observing a deviation from the analytic variance when applying the galactic mask of $f_{sky} = 20\%$.

\subsection{CB cross-correlation}\label{subsec:cross_corr}
\begin{figure}[ht]  
\includegraphics[width=.49\linewidth]{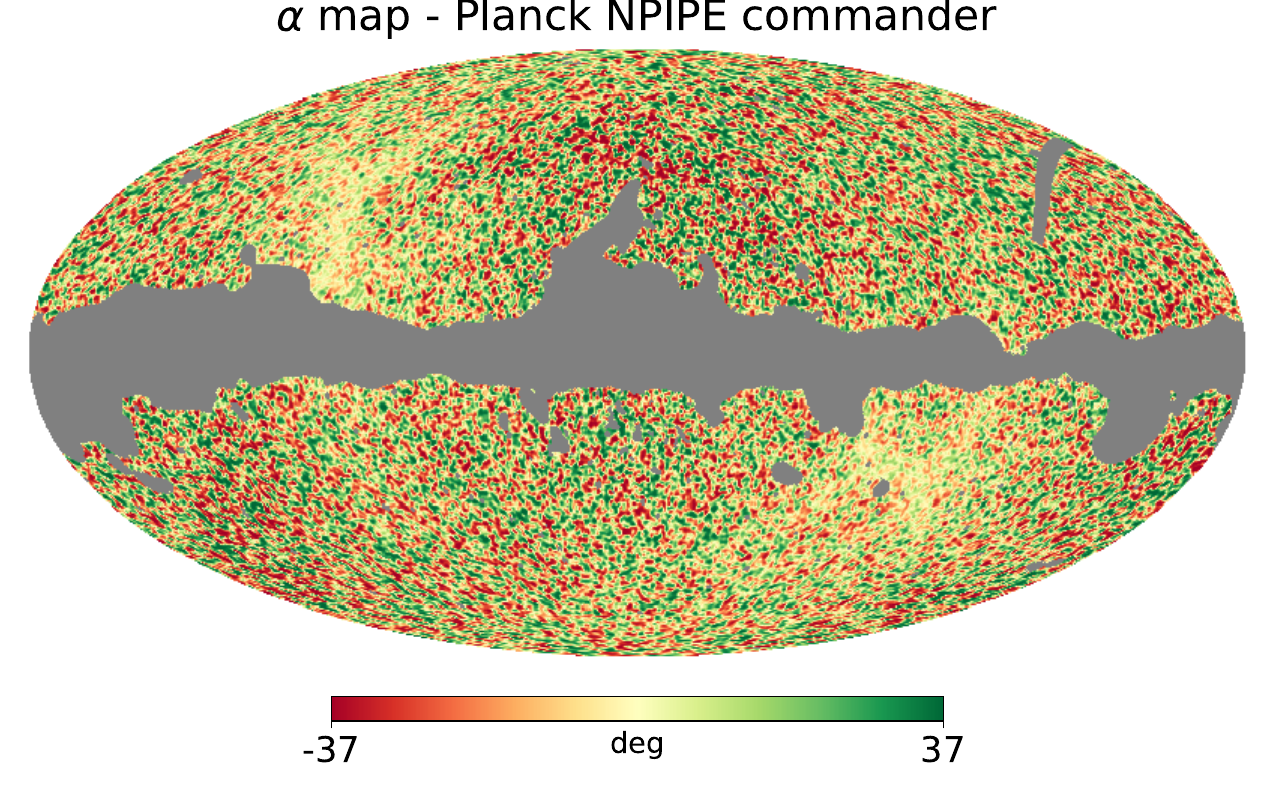}
\includegraphics[width=.49\linewidth]{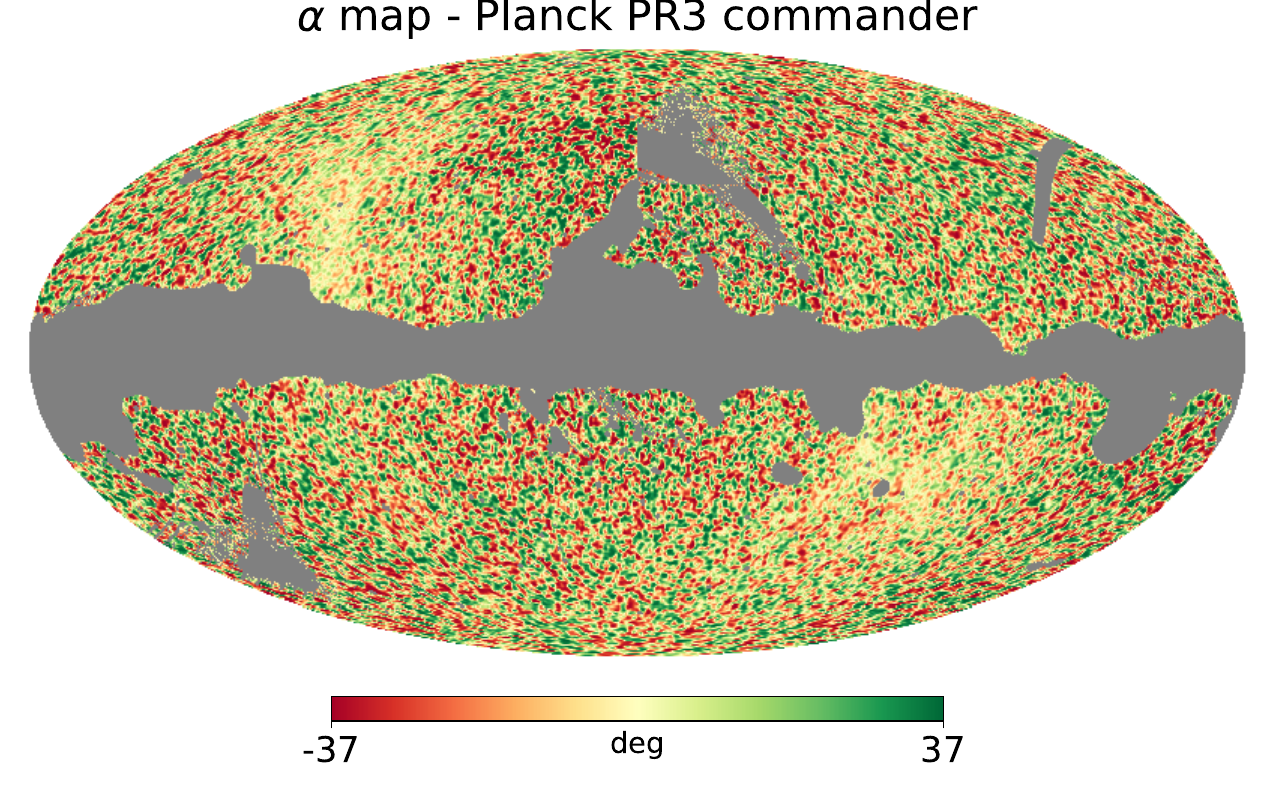}  
\caption{\label{fig:alpha_maps}CB map with a $1 deg$ smoothing evaluated from \planck\,NPIPE data products on the left, and the one evaluated from \planck\,PR3 data products on the right.}
\end{figure}

\begin{figure}[ht]
  \centering
  \begin{minipage}{\linewidth}
    \centering
    \includegraphics[width=.49\linewidth]{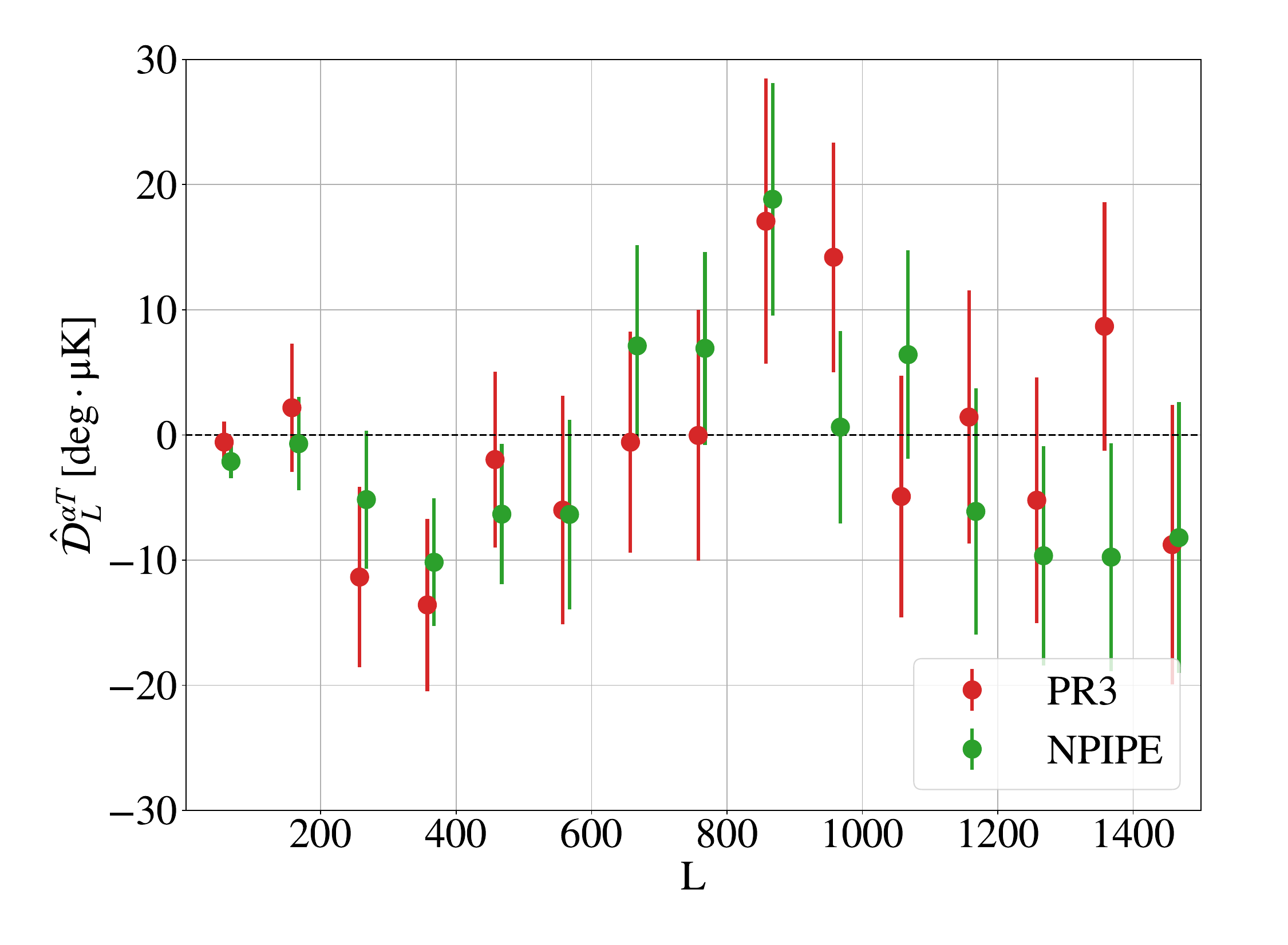}
  \end{minipage}
  \begin{minipage}{0.49\linewidth}
    \includegraphics[width=\linewidth]{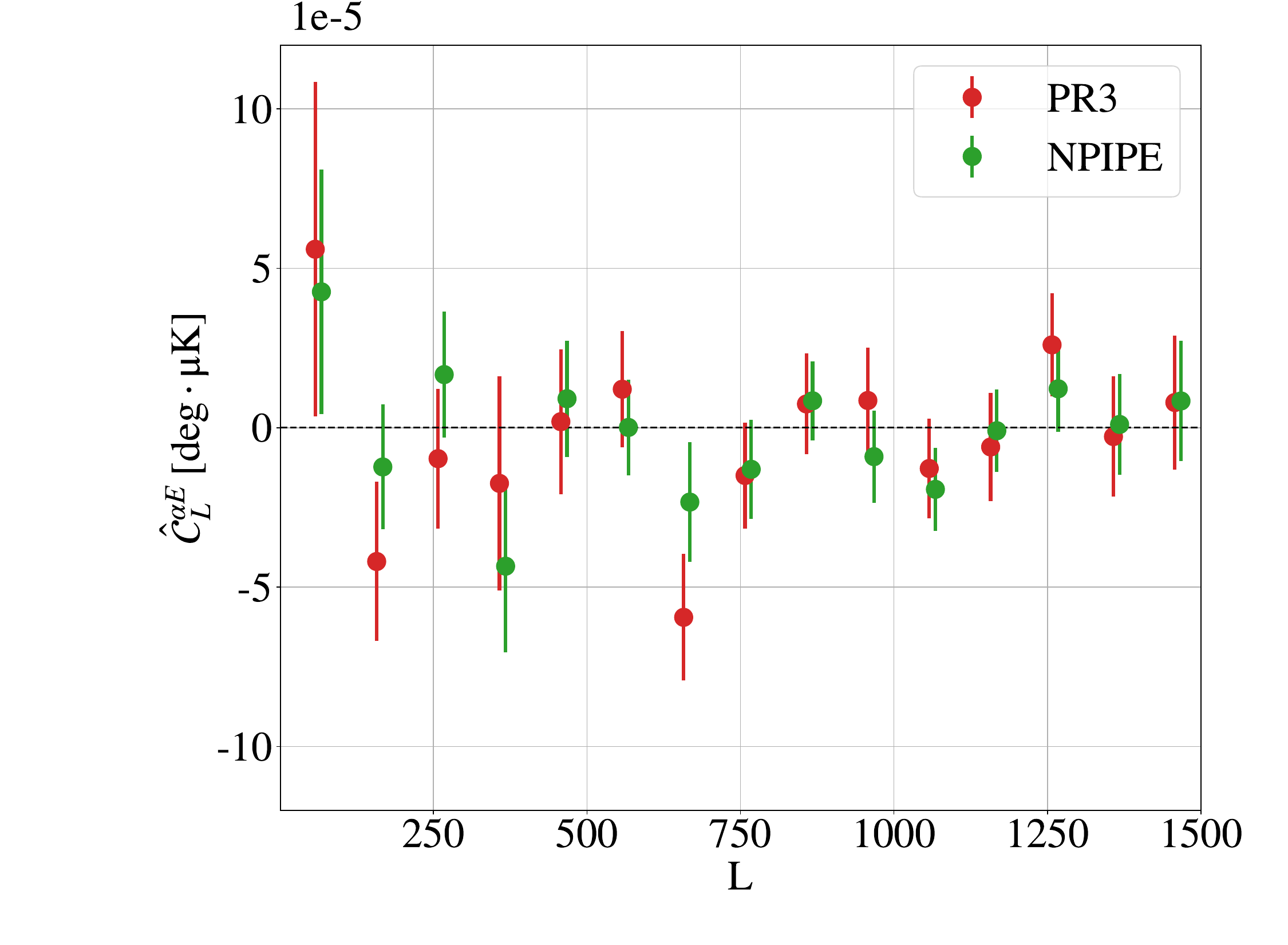}
  \end{minipage}
  \hfill
  \begin{minipage}{0.49\linewidth}
    \includegraphics[width=\linewidth]{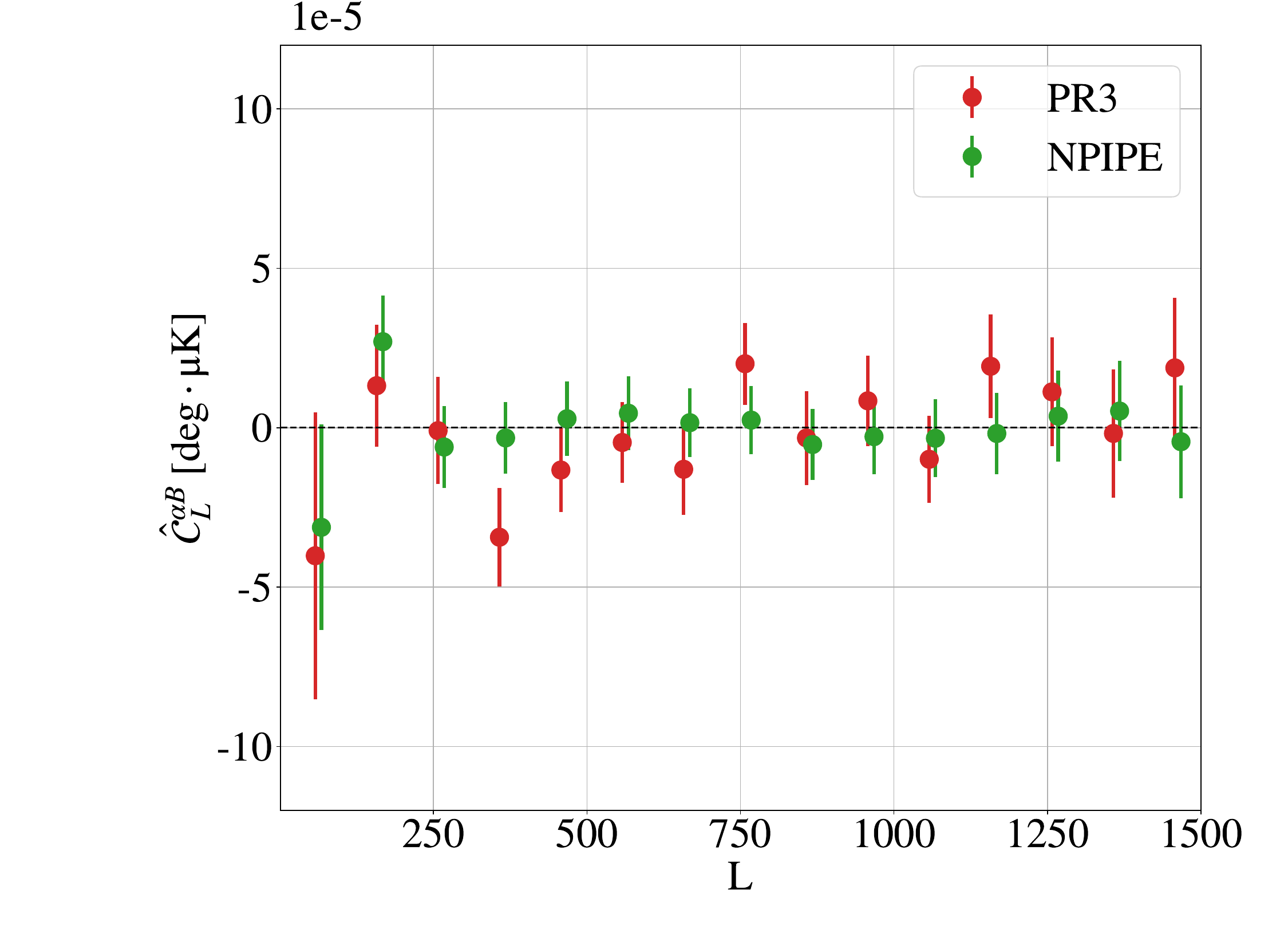}
  \end{minipage}
 \caption{\label{fig:cross-spectra}Cross-spectra between the CMB fields (temperature and polarization) and the CB field for NPIPE, in green, and PR3, in red, data products. \underline{Upper panel} $\alpha T$ cross-spectrum in \textit{band-powers}; \underline{Lower left panel} $\alpha E$ cross-spectrum; \underline{Lower right panel} $\alpha B$ cross-spectrum.}
\end{figure}

In this section, we present the CB field map, derived using the de-biased estimates of its spherical harmonic coefficients (eq. \eqref{de-biased alms}). This map is illustrated in figure \ref{fig:alpha_maps}. We then delve into the cross-correlation analysis between this field and the CMB temperature and polarization fields. Figures \ref{fig:cross-spectra} detail these results, showcasing data from both the \planck\,NPIPE (in green) and PR3 (in red) datasets.\par
The CB maps evaluated from \planck\,NPIPE (left panel in figure \ref{fig:alpha_maps}) and PR3 (right panel in figure \ref{fig:alpha_maps}) data products are processed using a $1^\circ$ FWHM Gaussian beam smoothing. This processing follows the subtraction of the mean field term from the $\alpha_{LM}$ estimates, based on both \planck\,NPIPE (map on the left) and PR3 (map on the right) data. The final maps have been masked using the standard masks of \planck\,NPIPE (first mask in figure \ref{fig:pol_mask}) and \planck\,PR3 (second mask in figure \ref{fig:pol_mask}). A notable feature in both CB maps is the correlation of smaller angle multipoles (yellow regions) with the \planck\,scanning strategy, as documented in \cite{planck2013-p03f}.\par
For the analysis of the cross-correlation between CB and CMB temperature and polarization fields, we employed the Pymaster Python package, which facilitated the calculation of the cross-spectra. In figure \ref{fig:cross-spectra}, the upper panel illustrates the cross-correlation between CB and the CMB temperature field ($\alpha T$) displayed in band-powers, for both \planck\,NPIPE (green points) and PR3 (red points) datasets. The lower panels display the cross-correlation of the CB field with the CMB polarization fields ($\alpha E$ and $\alpha B$). The power spectra displayed in figure \ref{fig:cross-spectra} are binned with 100 multipoles per bin, excluding the first 8 multipoles and the error bars have been evaluated from simulations. Note that the $\alpha T$ power spectrum is not represented in the units used for $\alpha E$ and $\alpha B$, rather it is showed in band-powers to allow for a more direct comparison with the $\alpha T$ power spectrum presented in \cite{SPT:2020cxx}.\par
Table \ref{tab: PTEs cross} presents the Probability To Exceed (PTE) values for the cross-correlation power spectra between the CB and CMB fields. Notably, the $\alpha B$ power spectrum from NPIPE, illustrated in the lower right panel of Figure \ref{fig:cross-spectra}, exhibits minimal scatter and this is reflected in its high PTE value ($98.75 \%$).\par
\begin{table}[ht]
\centering
\renewcommand{\arraystretch}{1.5}
\begin{tabular}{| c || c | c |}
 \hline
  & NPIPE \commander & PR3 \commander \\[1ex] 
  \hhline{|=||=|=|} 
 $\alpha T$ & $8.27 \%$ & $18.71 \%$\\[1ex]
 $\alpha E$ & $79.37 \%$ & $16.31 \%$\\[1ex]
 $\alpha B$ & $98.75 \%$ & $56.15 \%$\\[1ex]
 \hline
\end{tabular}
\caption{Probability To Exceed for the cross-spectra of the CB field with the temperature and polarization CMB fields.}
\label{tab: PTEs cross}
\end{table}

\section{Sensitivities of future experiments}\label{sec:forecasts}
In this section we present forecasts for forthcoming CMB experiments. The numerical computation of the variance of the estimator is crucial. Not only because it enables the calculation of the $\alpha_{LM}$ coefficients, but also serves as a rapid tool for forecasting future CMB experiments. By assessing $\sigma_L$ (eq. \eqref{sigma}), which reflects the sensitivity of an experiment to specific cross-correlations, we can identify the most promising avenues for detecting CB signatures.\par
We present forecasts for the \LB\,satellite \cite{LiteBIRD:2022cnt}, the \SO\,\cite{SimonsObservatory:2018koc} and \Sf\,\cite{Abazajian:2019eic}. Together with the sensitivities of these forthcoming CMB experiments, we also plot the one associated to the \planck\,\, satellite in order to provide a straightforward comparison. In the following, we briefly list the instrumental specifications used for the evaluation of the $\sigma_L$ for each experiment.\par
The \LB\,satellite, whose launch is predicted for the late 2020s, is composed of three telescopes which cover a total frequency range of $34$ - $448$ GHz. Its expected total sensitivity is $\sim 2.2\mu K\cdot arcmin$ with an angular resolution of 30 \textit{arcmin}. The ground-based \SO\,will cover the frequency range 27-280 GHz and is composed by the \texttt{Small Aperture Telescopes (SATs)} and the \texttt{Large Aperture Telescope (LAT)}. In this work we focus on the \texttt{Large Angular Telescope} \cite{Zhu_2021}. \SO\,\texttt{LAT} is expected to have a total sensitivity of $6 \mu K\cdot arcmin$ and an angular resolution of 1.4 \textit{arcmin}. \Sf\ is composed of 21 telescopes, 3 are large aperture telescopes and 18 are small aperture telescopes, and it will cover the frequency range 30-300 GHz. Its predicted total sensitivity is of $3 \mu K\cdot arcmin$ with an angular resolution of 1 \textit{arcmin}.
All the instrumental specifications, relevant for the analysis presented here, are summarised in table \ref{tab: CMB_exp}.
\begin{table}[ht]
\centering
\renewcommand{\arraystretch}{1.2}
\begin{tabular}{ |c||c|c|c|  }
 \hline
 \multicolumn{4}{|c|}{CMB experiments specifications} \\
 \hline
   & $\sigma_T$ & $\sigma_P $ & $\theta_{fwhm}$\\
   Experiment & $(\mu K arcmin)$ & $(\mu K arcmin)$ & (\textit{arcmin})\\
 \hline
 \planck\,& 40 & 56.57 & 5\\
 \LB\,& 2.2 & 3.26 & 30\\
 \SO\,\texttt{LAT} & 6 & 8.49 & 1.4\\
 \Sf\, & 3 & 4.24 & 1\\
 \hline
\end{tabular}
\caption{Instrumental specifications of the considered CMB experiments.}
\label{tab: CMB_exp}
\end{table}
\begin{figure}[ht]
\centering
\includegraphics[width=0.7\textwidth]{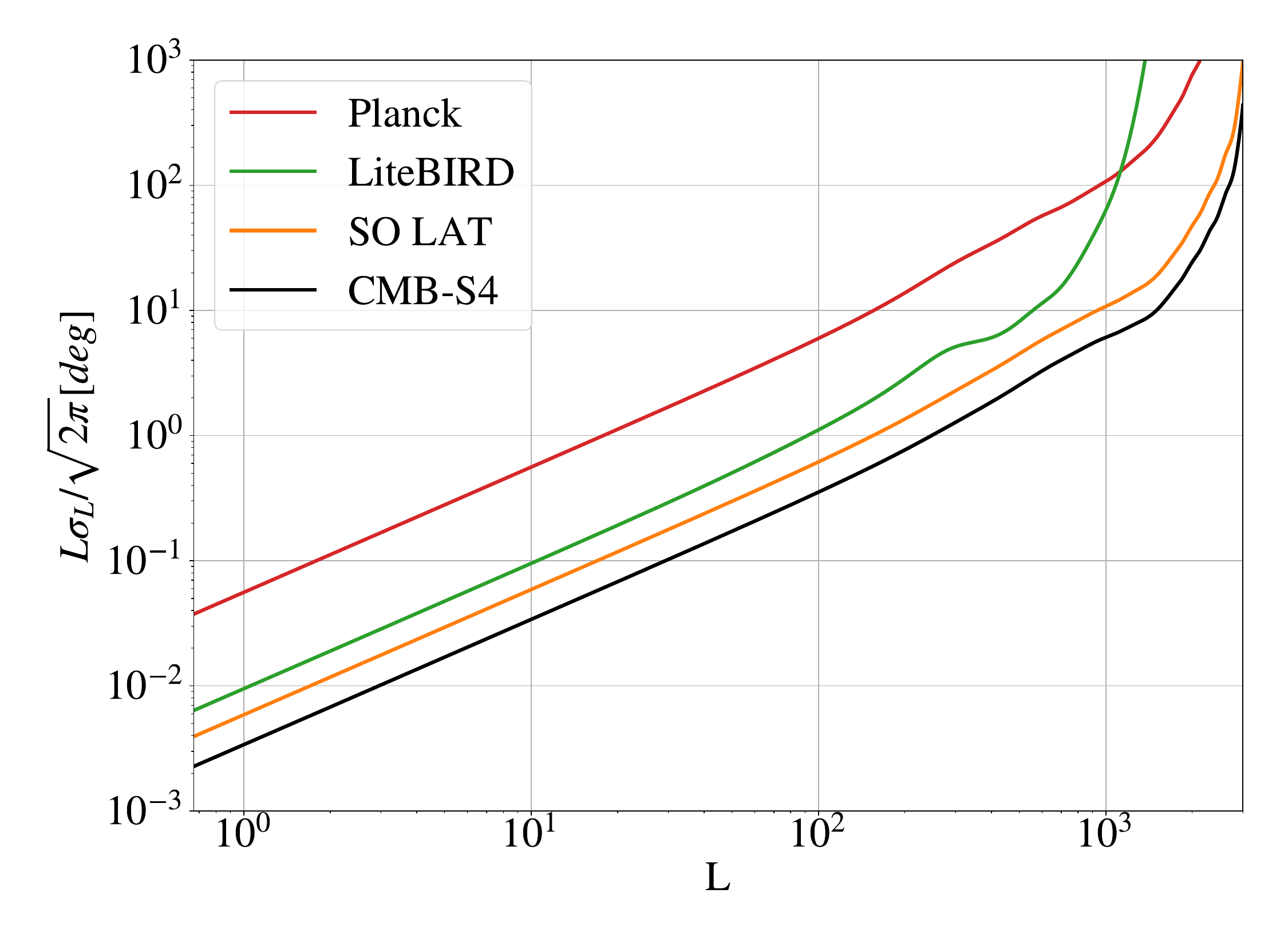}
\caption{\label{fig:sens_exp} Variance of the estimator evaluated with the instrumental specifications of the four CMB experiments reported in table \ref{tab: CMB_exp}.}
\end{figure}

Figure \ref{fig:sens_exp} shows the improvement of the \LB\,satellite with respect to \planck\,. Further improvements will be granted by ground based experiments. It is worth also to notice that the sensitivity that characterizes each experiment is not the only factor entering in the computation of $\sigma_L$, but also the angular resolution plays an important role. This can be seen comparing \LB\,and \SO\,\texttt{LAT} (as well as \LB\,and \Sf\,). Of course, experiments with higher resolution will grant access to smaller angular scales.\par

\section{Conclusions}\label{sec:conclusions}
In this study we have built an estimator for the spherical harmonic coefficients of the CB field which implements the method described in \cite{Gluscevic:2009mm}
and exploits the information contained in the CMB EB cross-correlation.
Based on the latter, we have built a pipeline aimed at extracting the angular power spectrum of CB from CMB polarization maps. We applied it to both \planck\,PR3 and NPIPE data products, considering \textit{full-mission} data and different data splits. In all cases, our analysis consistently found that the CB power spectrum is compatible with zero at a significance level of approximately $2\sigma$. As expected, \planck\,NPIPE \textit{full mission} data are slightly more constraining than the corresponding PR3 products.

Our findings agree with the CB power spectrum estimated from other analysis of \planck\, data \cite{Gruppuso:2020kfy, Bortolami:2022whx, Contreras:2017sgi} and various other experiments, including ACT \cite{Namikawa:2020ffr}, POLARBEAR \cite{POLARBEAR:2015ktq}, BICEP2/Keck Array \cite{BICEP:2021xfz} and SPT \cite{SPT:2020cxx}.
Moreover, we carried out a series of consistency checks reinforcing the reliability of our analysis, which we showed to be robust against: 1) the different component separation methods considered; 2) the different choices of minimum and maximum CMB multipoles included in the analysis; 3) the different masks applied to CMB maps.

Additionally, we employed the spherical harmonic coefficients of the CB field estimated from our pipeline to cross-correlate with the CMB temperature and polarization fields, producing $\alpha T$, $\alpha E$, and $\alpha B$ power spectra up to $L=1500$. Of these, the first two are predicted to be non-null in several models providing a further mean to constrain axion parameters, see e.g. \cite{Greco:2022ufo, Greco:2022xwj}.

We have also presented forecast for future CMB experiments showing that they will achieve sensitivities to anisotropic CB order of magnitudes better than what is currently available. 
In particular, the \LB\, satellite will reach a factor of $\sim 25$ improvement with respect to the \planck\,at power spectrum level, while \Sf\ \cite{Abazajian:2019eic} will be able to reach an improvement of a factor of $\sim 1000$.

The code and the pipeline developed for this work are publicly available\footnote{The code is available on GitHub at \url{https://github.com/paganol/alpha_lm}}, along with products employed\footnote{CB spectra from \planck\,PR3 and NPIPE, as well as its cross-correlations with the CMB T-, E- and B-fields are available on GitHub at \url{https://github.com/giorgiazagatti/CB_Planck_maps_spectra.git}}. 
Upon requests we can provide additional products such as birefringence maps or $\alpha_{LM}$ coefficients.

\acknowledgments
We thank Margherita Lembo for useful discussions on the implementation of the estimator. We acknowledge the financial support from the INFN InDark initiative and from the COSMOS network (www.cosmosnet.it) through the ASI (Italian Space Agency) Grants 2016-24-H.0 and 2016-24-H.1-2018, as well as 2020-9-HH.0 (participation in \LB\,phase A). GF acknowledges the support of the European Research Council under the Marie Sk\l{}odowska Curie actions through the Individual Global Fellowship No.~892401 PiCOGAMBAS.
This work is supported in part by the MUR PRIN2022 Project “BROWSEPOL: Beyond standaRd mOdel With coSmic microwavE background POLarization”-2022EJNZ53 financed by the European Union - Next Generation EU.
We acknowledge the use of \texttt{numpy} \citep{Harris:2020xlr} and \texttt{matplotlib} \citep{Hunter:2007ouj} software packages, and the use of computing facilities at CINECA.
Some of the results in this paper have been derived using the \texttt{healpy} \citep{Zonca:2019vzt} and \healpix\ \citep{Gorski:2004by} packages.

\bibliographystyle{JHEP}
\bibliography{bibliography,Planck_bib}

\appendix
\section{Construction of the estimator} \label{app:impl est}
In section \ref{subsec:eb-estimator} we re-write the part of equation \eqref{alpha_est} without its normalization (i.e., without $\sigma_L^{-2}$), that we indicate as $\overline{\alpha}_{LM}^{UN}$, in order to reduce the computation time.
\begin{eqnarray}\label{num_alpha}
    \overline{\alpha}_{LM}^{UN} = \displaystyle\sum_{\ell\geq\ell'}(1+\delta_{\ell\ell'})^{-1}&\left\{\dfrac{F_{\ell\ell'}^{L,EB}\displaystyle\sum_{mm'}a_{\ell m}^{E,map}a_{\ell'm'}^{B,map,*}\xi_{\ell m\ell'm'}^{LM}}{C_\ell^{EE,map}C_{\ell'}^{BB,map}} \right. + \nonumber\\
    & + \left. \dfrac{F_{\ell\ell'}^{L,BE}\displaystyle\sum_{mm'}a_{\ell m}^{B,map}a_{\ell'm'}^{E,map,*}\xi_{\ell m\ell'm'}^{LM}}{C_\ell^{BB,map}C_{\ell'}^{EE,map}}\right\} .
\end{eqnarray}
In this appendix we show the calculations to obtain the implemented expression of the harmonic estimator for the CB field.\par
Since the second term inside the summation is equal to the first one with $\ell-\ell'$ inverted, an equivalent expression of \eqref{num_alpha}, making the expression of the $F_{\ell\ell'}^{L,EB}$ term explicit, is:
\begin{equation}
    \overline{\alpha}_{LM}^{UN}= \displaystyle\sum_{\ell\ell'}\dfrac{2H_{\ell\ell'}^L C_\ell^{EE}W_\ell W_{\ell'}\displaystyle\sum_{mm'}a_{\ell m}^{E,map}a_{\ell'm'}^{B,map,*}\xi_{\ell m\ell'm'}^{LM}}{C_\ell^{EE,map}C_{\ell'}^{BB,map}} .
\end{equation}
In the following, we re-write the $2H_{\ell\ell'}^L \xi_{\ell m\ell'm'}^{LM}$ term. Specifying the expression of $\xi_{\ell m\ell'm'}^{LM}$ we end up with:
\begin{equation}\label{prod}
    2H_{\ell\ell'}^L\sqrt{\dfrac{(2\ell+1)(2L+1)(2\ell'+1)}{4\pi}}\begin{pmatrix}\ell & L & \ell' \\ -m & M & m'\end{pmatrix} ,
\end{equation}
and, exploiting the definition of $H_{\ell\ell'}^L$ and the properties of the Wigner-3j symbols, recalling that $\ell+L+\ell'$ must be even in the case of the EB cross-correlation induced by CB, it is also possible to write:
\begin{equation*}
    2H_{\ell\ell'}^L = \begin{pmatrix}
        \ell & L & \ell' \\
        +2 & 0 & -2
    \end{pmatrix}+\begin{pmatrix}
        \ell & L & \ell' \\
        -2 & 0 & +2
    \end{pmatrix} ,
\end{equation*}
so that we can use the definition of the triple integral to re-write the product \eqref{prod} as:
\begin{multline}
    \sqrt{\dfrac{(2\ell+1)(2L+1)(2\ell'+1)}{4\pi}}\Biggl[\begin{pmatrix}
        \ell & L & \ell' \\
        +2 & 0 & -2
    \end{pmatrix}+\begin{pmatrix}
        \ell & L & \ell' \\
        -2 & 0 & +2
    \end{pmatrix} \Biggr]\begin{pmatrix}\ell & L & \ell' \\ -m & M & m'\end{pmatrix} = \\
    \int d\hat{n}Y_{LM}(\hat{n})[_{-2}Y_{\ell -m}(\hat{n})_{+2}Y_{\ell'm'}(\hat{n})+_{+2}Y_{\ell -m}(\hat{n})_{-2}Y_{\ell'm'}(\hat{n})] .
\end{multline}
Therefore, the estimator can be written as:
\begin{multline}\label{new_est_APP}
\overline{\alpha}_{LM}^{UN} = \int d\hat{n} Y_{LM} \Biggl[\displaystyle\sum_{\ell m}\dfrac{C_\ell^{EE}(-1)^m a_{\ell m}^{E,map}W_\ell {}_{-2}Y_{\ell -m}}{C_\ell^{EE,map}}\displaystyle\sum_{\ell'm'}\dfrac{a_{\ell'm'}^{B,map,*}W_{\ell'}{}_{+2}Y_{\ell'm'}}{C_{\ell'}^{BB,map}}+\\
\displaystyle\sum_{\ell m}\dfrac{C_\ell^{EE}(-1)^m a_{\ell m}^{E,map}W_\ell {}_{+2}Y_{\ell -m}}{C_\ell^{EE,map}}\displaystyle\sum_{\ell'm'}\dfrac{a_{\ell'm'}^{B,map,*}W_{\ell'}{}_{-2}Y_{\ell'm'}}{C_{\ell'}^{BB,map}}\Biggr] = \\
\int d\hat{n} Y_{LM} \Biggl[\displaystyle\sum_{\ell m}\dfrac{C_\ell^{EE}a_{\ell m}^{E,map,*}W_\ell {}_{-2}Y_{\ell m}}{C_\ell^{EE,map}}\displaystyle\sum_{\ell'm'}\dfrac{a_{\ell'm'}^{B,map,*}W_{\ell'}{}_{+2}Y_{\ell'm'}}{C_{\ell'}^{BB,map}}+\\
\displaystyle\sum_{\ell m}\dfrac{C_\ell^{EE}a_{\ell m}^{E,map,*}W_\ell {}_{+2}Y_{\ell m}}{C_\ell^{EE,map}}\displaystyle\sum_{\ell'm'}\dfrac{a_{\ell'm'}^{B,map,*}W_{\ell'}{}_{-2}Y_{\ell'm'}}{C_{\ell'}^{BB,map}}\Biggr] ,
\end{multline}
where the second equivalence follows from $(a_{\ell m}^X)^*=(-1)^m a_{\ell -m}^X$.\\
At this point we define two new objects:
\begin{align}\label{QUobjects_E_APP}
    Q^E \pm iU^E = \displaystyle\sum_{\ell m}(C_\ell^{EE}\overline{a}_{\ell m}^{E,*}){}_{\pm 2}Y_{\ell m}, \\\label{QUobjects_B_APP}
    Q^B \pm iU^B = \displaystyle\sum_{\ell m}(\pm i\overline{a}_{\ell m}^{B,*}){}_{\pm 2}Y_{\ell m} ,
\end{align}
with $\overline{a}_{\ell m}^{E,*}$ and $\overline{a}_{\ell m}^{B,*}$ defined as:
\begin{align}
    \overline{a}_{\ell m}^{E,*} = \dfrac{a_{\ell m}^{E,map,*}}{C_\ell^{EE,map}}W_\ell , \\
    \overline{a}_{\ell m}^{B,*} = \dfrac{a_{\ell m}^{B,map,*}}{C_\ell^{BB,map}}W_\ell .
\end{align}
Re-writing equation \eqref{new_est_APP} in terms of \eqref{QUobjects_E_APP} and \eqref{QUobjects_B_APP}, we obtain:
\begin{equation}
    \overline{\alpha}_{LM}^{UN} = \int d\hat{n} Y_{LM}[2(Q^EU^B-U^EQ^B)].
\end{equation}
Performing the complex conjugate of the above equation and recognizing a "map-like" object in the term inside the square brackets, we obtain:
\begin{equation}
    \overline{\alpha}_{LM}^{UN,*} = \int d\hat{n}Y_{LM}^* m'(\alpha) .
\end{equation}
This final expression is particularly useful since, having defined $m'(\alpha)$ in terms of the the real Q-like and U-like objects from \eqref{QUobjects_E_APP} and \eqref{QUobjects_B_APP}, we can directly obtain the associated spherical harmonic coefficients.\\
A word of caution before proceeding. This procedure allows us to obtain an estimate for the \textit{complex conjugate} and \textit{unnormalized} $\alpha_{LM}$ coefficients. We recover the final estimate for the $\alpha_{LM}$ coefficients as:
\begin{equation}
    \overline{\alpha}_{LM} = \dfrac{(\overline{\alpha}_{LM}^{UN,*})^*}{\sigma_L^{-2}} .
\end{equation}

\section{Validation tests} \label{app:validation}
In this appendix we show some validation tests for our pipeline. In particular, we show that the application of the implemented estimator to a rotated CMB polarization map recovers the same input rotation. On the contrary, we also test the de-biasing procedure, showing that the pipeline produces a CB power spectrum compatible with zero in case of un-rotated input CMB polarization maps.\par
All the presented results have been obtained using a set of 100 CMB full-sky simulations at a resolution of \Nside = 512, with $\ell_{max}^{CMB} = 512$, $\theta_{fwhm} = 30\ arcmin$, $\sigma_{noise}^T = 1\mu K arcmin$ and $\sigma_{noise}^P = \sqrt{2}\sigma_{noise}^T$.

\subsection{Validation with a rotation signal}

In order to validate the pipeline in case of a non-zero rotation signal in the input CMB polarization maps, we use a CB power spectrum that is scale invariant in band powers:
\begin{equation*}
    D^{\alpha\alpha}_L = \mathcal{A} = \dfrac{L(L+1)}{2\pi}C^{\alpha\alpha}_L .
\end{equation*}
After a proper choice of the amplitude, $\mathcal{A}=0.006\ deg^2$, we can end up with the fiducial CB power spectrum used for the test of the estimator, by inverting the previous relation:
\begin{equation}\label{Cl}
    C^{\alpha\alpha, fid}_L = \dfrac{2\pi\mathcal{A}}{L(L+1)} . 
\end{equation}
At this point we generate 100 realizations of CMB maps from the fiducial power spectrum and we rotate each CMB realization accordingly to the associated CB field realization. We are generating different CB realizations for each simulation, each one obtained from the same input $\alpha\alpha$ power spectrum. Once we have the rotated CMB maps, we convolve for the beam and add the noise to each realization.\par
Having the set of rotated CMB+noise simulations, we apply the pipeline described in this work to the simulated maps, evaluating the $\alpha\alpha$ power spectrum before the de-biasing procedure, $C_L^{\hat{\alpha}\hat{\alpha}}$, and the isotropic bias term, $C_L^{bias,iso}$. Since for the validation part we are working in the ideal case of full-sky and white noise, only the computation of the isotropic bias term is needed.\par
\begin{figure}[ht]
\centering
\includegraphics[width=0.7\textwidth]{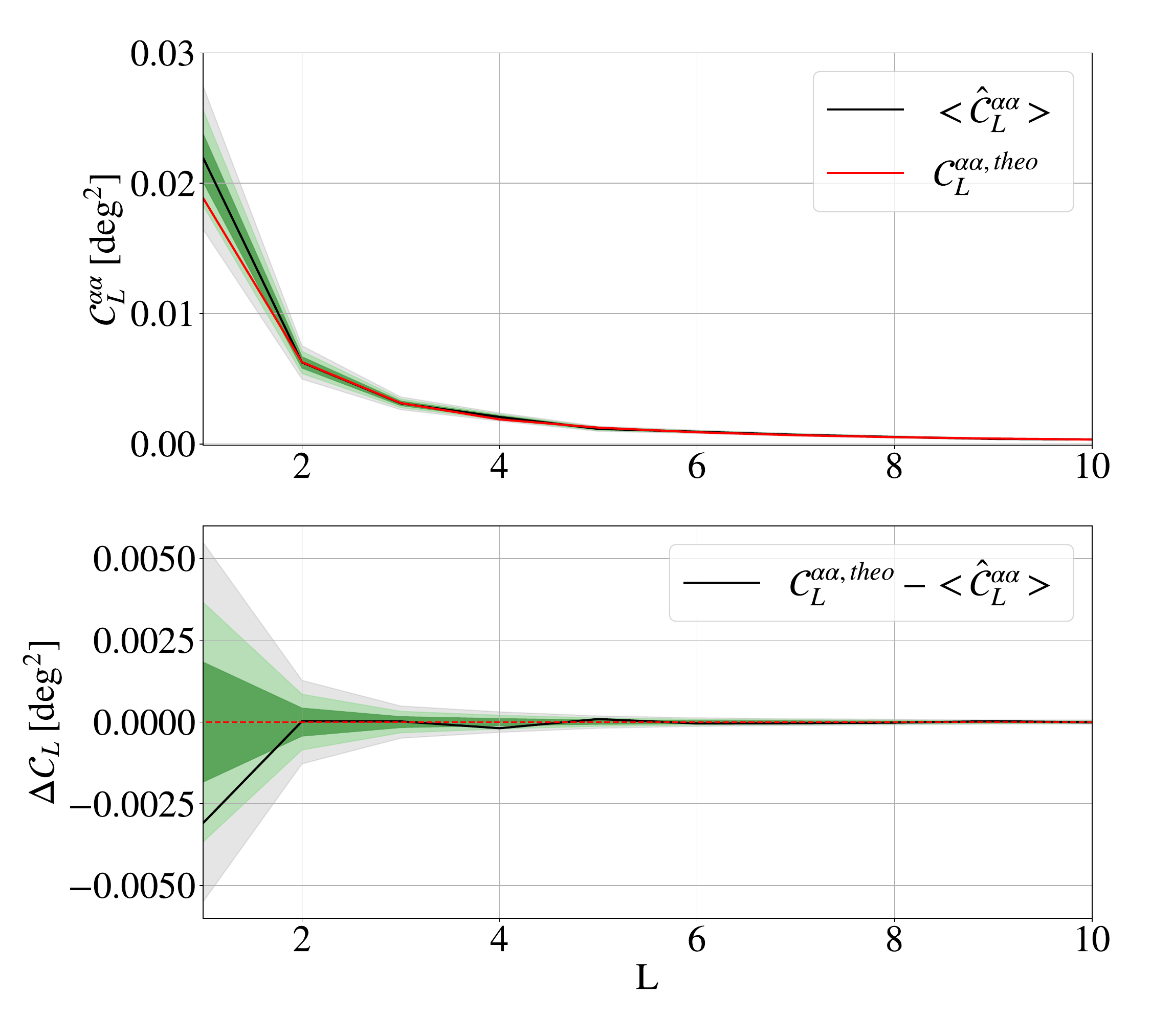}
\caption{\label{fig:val rot}Validation with rotation signal: \underline{Upper panel} Input power spectrum (\textit{red curve}), average of the estimated $\alpha\alpha$ power spectra (\textit{black curve}), the $1 \sigma$, $2 \sigma$ and $3 \sigma$ confidence intervals are the \textit{dark green}, \textit{light green} and \textit{gray} areas, respectively. \underline{Lower panel} Difference between the input and the recovered power spectrum.}
\end{figure}
In the upper panel of figure \ref{fig:val rot} we compare the input signal and the average of the estimated CB power spectra computed over the simulations. In the lower panel we show the difference between the estimated and the input power spectra. The difference is compatible with zero at $2 \sigma$.

\subsection{Validation without a rotation signal}

For this second part of the validation we apply our pipeline to the 100 full-sky CMB polarization maps without rotating them.\par
\begin{figure}[ht]
\centering
\includegraphics[width=0.7\textwidth]{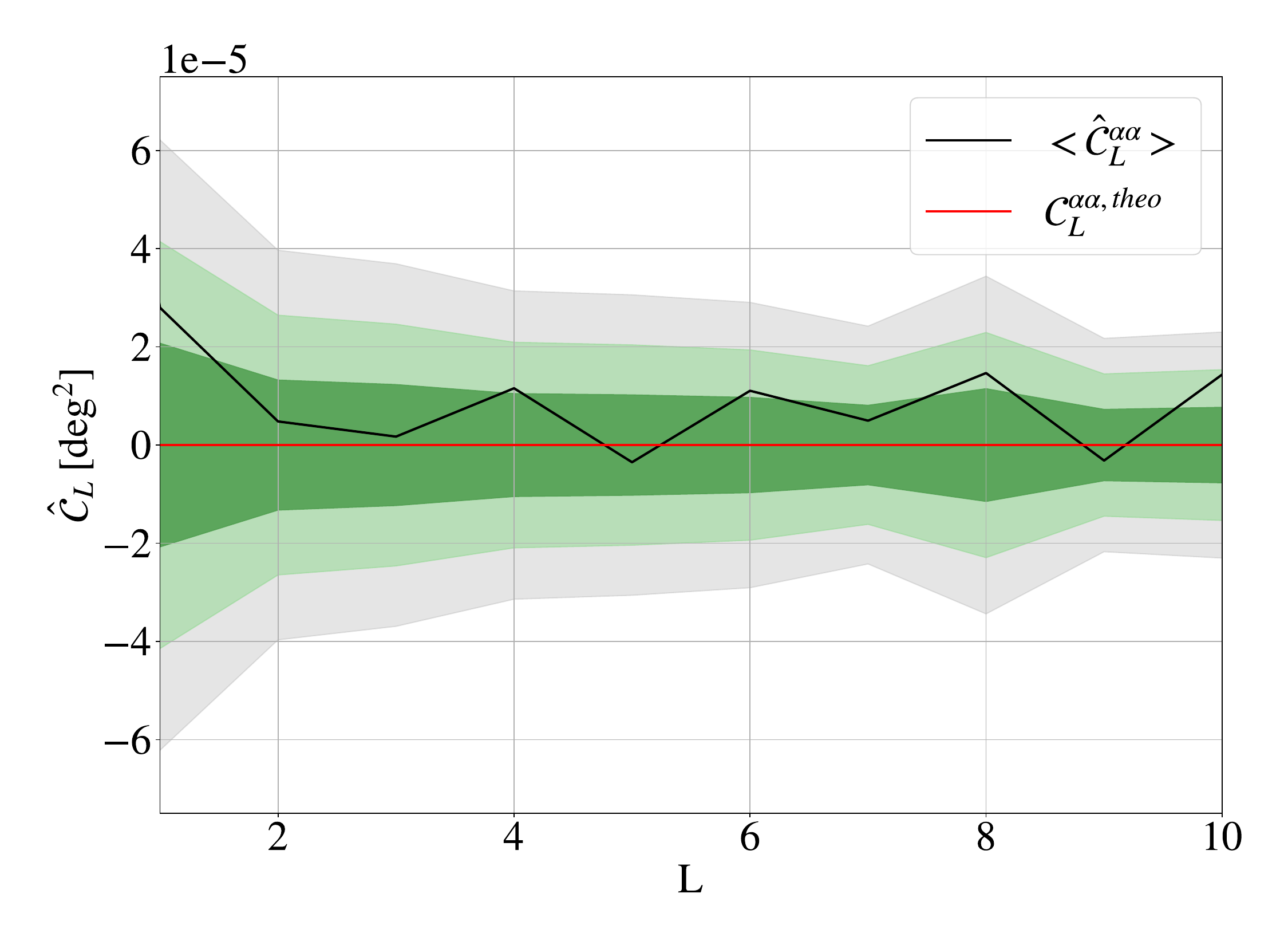}
\caption{\label{fig:val no rot}Average of the estimated $\alpha\alpha$ power spectra (\textit{black curve}) from simulated CMB realizations without rotation. The $1 \sigma$, $2 \sigma$ and $3 \sigma$ confidence intervals are the \textit{dark green}, \textit{light green} and \textit{gray} areas, respectively.}
\end{figure}
In figure \ref{fig:val no rot} we compare the average of the $\alpha\alpha$ power spectra computed over the simulations (\textit{black curve}) with the expected zero rotation signal (\textit{red curve}). The shaded areas are the $1\sigma$, $2\sigma$ and $3\sigma$ confidence intervals. The average de-biased power spectrum is compatible with zero at $2\sigma$.
\end{document}